\documentclass{article}
\usepackage{graphicx} 
\usepackage{amsmath,amssymb,bm,mathtools,dsfont,upgreek,stmaryrd,kpfonts}
\usepackage{authblk}
\usepackage{biblatex}
\usepackage{xcolor}
\newtheorem{theorem}{Theorem}[section]
\newtheorem{assumptions}[theorem]{Assumptions}
\newtheorem{definition}[theorem]{Definition}
\newtheorem{pro}[theorem]{Proposition}
\newtheorem{thm}[theorem]{Theorem}
\newtheorem{lemma}[theorem]{Lemma}
\newtheorem{cor}[theorem]{Corollary}

\newcommand{\proof}{\noindent {\it Proof. }}
\newcommand{\qed}{\vspace{.2cm}~\hfill{$\square$}\newline\noindent}
\newcommand{\QED}{~\hfill{$\square$}}

\newcommand{\norm}[1]{\ensuremath{\lVert #1 \rVert}}
\newcommand{\abs}[1]{\ensuremath{\lvert #1 \rvert}}

\title{Periodicity of atomic structure \break in a Thomas-Fermi mean-field model}
\author[1]{August Bjerg}
\author[1]{Jan Philip Solovej}
\affil[1]{\centering QMATH, Department of Mathematical Sciences, \break University of Copenhagen, Universitetsparken 5, \break DK-2100, Copenhagen Ø, Denmark}
\date{}

\begin{document}

\maketitle

\begin{abstract}
\noindent We consider a Thomas-Fermi mean-field model for large neutral atoms. That is, Schrödinger operators $H_Z^{\text{TF}}=-\Delta-\Phi_Z^{\text{TF}}$ in three-dimensional space, where $Z$ is the nuclear charge of the atom and $\Phi_Z^{\text{TF}}$ is a mean-field potential coming from the Thomas-Fermi density functional theory for atoms. For any sequence $Z_n\to\infty$ we prove that the corresponding sequence $H_{Z_n}^{\text{TF}}$ is convergent in the strong resolvent sense if and only if $D_{\text{cl}}Z_n^{1/3}$ is convergent modulo $1$ for a universal constant $D_{\text{cl}}$. This can be interpreted in terms of periodicity of large atoms. We also characterize the possible limiting operators (infinite atoms) as a periodic one-parameter family of self-adjoint extensions of $-\Delta-C_\infty\abs{x}^{-4}$ for an explicit number $C_\infty$. 
\end{abstract}

\tableofcontents

\section{Introduction}
The motivation for the mathematical work in this paper is to understand the periodicity of the periodic table of the elements. More precisely, the question is why atoms in the groups of the periodic table, e.g., the noble gases or the alkali atoms, have very similar chemical properties.

Before we can properly ask this question we must first understand what even defines the different groups. From elementary chemistry  we know that this is related to filling electrons in atomic orbitals that span the subspaces with angular momentum quantum numbers $\ell=0,1,2,\dots$.
The alkali atoms are those atoms where a new $\ell=0$ orbital is being occupied by an electron (referred to as an $s$-electron in chemistry). The noble gases are those atoms where all\footnote{The factor $2$ in $2(2\ell+1)$ is counting spin.} $2(2\ell+1)=6$ electrons in an $\ell=1$ subspace have been filled ($p$-electrons in the chemist's notation). A natural question is of course in which order the different $\ell$ subspaces are being filled. In chemistry this is described by the empirical Aufbau principle (or Madelung rule \cite{Madelung}). We shall not describe the rule in details here, but note that it gives us a general formula for the atomic number $Z_\ell(n)$, where we start filling an $\ell$ subspace for the $n^{\text{th}}$ time. This general formula is 
\begin{equation}\label{ZellFormula}
Z_{\ell}(n)=\frac{(n+2\ell-1)((n+2\ell)^2+4(n+2\ell)+9)}6-\frac{(1+(-1)^{n})(n+2\ell+1)}4+1-2\ell(\ell+2).
\end{equation}
This formula is indeed reflected in the periodic table and is correct for the atoms
\[
    \begin{tabular}{rl}
         $\ell=0$:& The alkali atoms in group $1$ where we start filling a new $s$ orbital, \\& 
          i.e., $Z=1,3,11,19,37,55, 87$ \\
         $\ell=1$:& Where we fill a new $p$ subspace, i.e.,  $Z=5,13,31,49,81,113$\\
         $\ell=2$:& Where we fill a new $d$ subspace, i.e., $Z=21,39,71$
    \end{tabular}
\]
It most likely fails, however, for the case $\ell=2$ and $n=4$. Here we have\footnote{This has been a somewhat contested issue, see \cite{Lawrencium}.}  $Z_2^{\rm true}(4)=104$, but the formula above gives $Z_2(4)=103$. Moreover, it fails generally for $\ell=3$ where $Z_3^{\rm true}(1)=58$ and $Z^{\rm true}_3(2)=91$, but the formula gives $Z_3(1)=57$ and $Z_3(2)=89$. There are several other exceptions in the periodic table to the general Madelung rule.  

Fermi \cite{FermiAufbau} attempted to calculate $Z_\ell(n)$  in a model where electrons move independently in a mean-field potential describing the effect of the interaction of all the other electrons. 
Fermi used the mean-field potential derived from his own Thomas-Fermi 
model \cite{TFFermi,TFThomas}. 
The formula he derived, however, does not agree with the above expression. In particular it does not reproduce the $1/6$ in the leading order term in $n$.

Other attempts \cite{KlechkovskiiAufbau,TietzAufbau, WongAufbau}, used a different mean-field potential suggested by Tietz in \cite{TietzAufbau} which does reproduce the $1/6$ asymptotically for large $n$. To the best of our knowledge there are no justifications for the use of the Tietz potential other than that it reproduces the Madelung rule (asymptotically). 

In the full many-body quantum mechanical description of atoms the concept of electron orbitals is not well-defined. A possible approach is to consider the natural orbitals, i.e., the eigenfunctions of the 1-particle reduced density matrix $\gamma$ of an atomic many-body ground state. These eigenfunctions are, however, unlikely to be labelled by angular momenta. We may of course always ask for the occupation number $n_\ell(Z)=\operatorname{Tr}[\gamma P_\ell]$ of the ground state in an angular momentum eigenspace given by the projection $P_\ell$. In a forthcoming publication\footnote{In preparation in joint work with S\o ren Fournais and Peter Hearnshaw.} we will show that $n_\ell(Z)$, as defined above, does not satisfy the Madelung rule asymptotically for large $Z$ and almost all $\ell$ in a sense to be made precise in the publication. In fact, it turns out that Fermi's formula gives the correct answer here. 

We will in the present paper consider exactly the same model as Fermi and will describe this model in more details in the next subsection. For each $Z$ it gives a spherically symmetric mean-field potential and a corresponding mean-field Schr\"odinger operator. Fermi's idea was to ask whether the ordering of the energy levels --- as a function of angular momenta --- of this mean-field operator agrees with experimental data. 

We will address the somewhat different issue of whether the model explains the similarity in chemical properties for certain sequences of atomic numbers, corresponding to the groups in the periodic table. To phrase this as a more mathematical question we ask whether the Thomas-Fermi mean-field operators converge in some appropriate sense as $Z$ tends to infinity through certain sequences. The main result of this paper (Theorem ~\ref{MainResultTF}) is that this is, indeed, the case in the sense of strong resolvent convergence of operators. Moreover, the sequences of $Z$ agree with what Fermi found in his attempt to explain the structure of the periodic table. Note that strong resolvent convergence implies that the spectrum of the limiting operator is included in the limits of the spectra (spectral exclusion). Since spectra describe chemical properties such as, e.g.,  the ionization energies we may interpret our result as saying that these sequences represent atoms with similar chemical properties in this model. The question whether ionization energies converge as $Z$ tends to infinity through particular sequences was studied numerically in several density functional theories in \cite{Constantin2010}, where it was found that this is, indeed, the case.  

\subsection{Thomas-Fermi theory for atoms}
Our mean-field model is based on the Thomas-Fermi density functional theory introduced in \cite{TFThomas,TFFermi}. We review now briefly some mathematical facts concerning this and refer to \cite{TFLiebSimon} or \cite{TFLieb} for further details.

We consider $3$-dimensional space. The energy of an atom with atomic number $Z$ and electron density $\rho$ is in Thomas-Fermi theory given by
\begin{equation}\label{TFFunctional}
\mathcal{E}_Z^{\text{TF}}[\rho]=c_{\text{TF}}\int\rho(x)^{5/3}\,dx-Z\int\frac{\rho(x)}{\abs{x}}\,dx+\frac{1}{2}\int\int\frac{\rho(x)\rho(y)}{\abs{x-y}}\,dx\,dy
\end{equation}
where $c_{\text{TF}}=\frac35(3\pi^2)^{2/3}$. We have here\footnote{These are the units used throughout the paper.} used the units that $\hbar=e=2m=1$, where $m$ is the electron mass, and consider the case with spin $1/2$, i.e. $2$ spin degrees of freedom. Thomas-Fermi theory can be modified to include any spin degree of freedom by changing only the value of $c_{\text{TF}}$. It is known that the infimum
\begin{equation}\label{MinimizationProblem}
\inf_{\substack{\rho\in L^1\cap L^{5/3}\\ \rho\geq0}}\mathcal{E}_Z^{\text{TF}}[\rho]
\end{equation}
is attained for some unique spherically symmetric $\rho_Z^{\text{TF}}$ which is smooth on $\mathbb{R}^3\backslash\{0\}$ and has total mass $Z$. Of great importance to us will be the quantity
\begin{equation}\label{TFPotential}
\Phi_Z^{\text{TF}}(x):=\frac{Z}{\abs{x}}-\int\frac{\rho_Z^{\text{TF}}(y)}{\abs{x-y}}\,dy
\end{equation}
called the Thomas-Fermi potential. This clearly inherits spherical symmetry and smoothness from $\rho_Z^{\text{TF}}$, and it is moreover strictly positive. It describes the electrostatic interactions between a fixed electron and all electrons in the atom (itself included). From the minimization problem (\ref{MinimizationProblem}) one additionally finds that
\begin{equation}\label{TFEquation}
\Phi_Z^{\text{TF}}=\frac{5c_{\text{TF}}}{3}(\rho_Z^{\text{TF}})^{2/3},\qquad\text{yielding}\qquad\Delta\Phi_Z^{\text{TF}}=4\uppi\;\Bigl(\frac{3\Phi_Z^{\text{TF}}}{5c_{\text{TF}}}\Bigr)^{3/2}.
\end{equation}
Here, the first equation is called the Thomas-Fermi equation, and the latter, valid on $\mathbb{R}^3\backslash\{0\}$, is obtained by combining this with the definition of $\Phi_Z^{\text{TF}}$. The fact that $\Phi_Z^{\text{TF}}$ satisfies this differential equation together with some qualitative observations can be used to prove the asymptotics
\begin{equation}\label{TFAsymp}
\Phi_1^{\text{TF}}(x)=\Bigl(\frac{5c_{\text{TF}}}{3}\Bigr)^3\frac{9}{\uppi^2\abs{x}^4}+o_{\abs{x}\to\infty}(\abs{x}^{-4})
\end{equation}
for the Thomas-Fermi potential near infinity. Moreover, it can be easily deduced from (\ref{TFPotential}) that $\abs{x}\Phi_Z^{\text{TF}}(x)\to Z$ as $\abs{x}\to0$.

We notice further that it follows directly from the definition (\ref{TFFunctional}) of the energy functional that $\mathcal{E}^{\text{TF}}_Z[Z^2\rho(Z^{1/3}\,\cdot\,)]=Z^{7/3}\mathcal{E}^{\text{TF}}_1[\rho]$. From this we learn that $\rho_Z^{\text{TF}}(x)=Z^2\rho_1^{\text{TF}}(Z^{1/3}x)$ and in turn, by (\ref{TFPotential}) or (\ref{TFEquation}), $\Phi_Z^{\text{TF}}(x)=Z^{4/3}\Phi_1^{\text{TF}}(Z^{1/3}x)$. These perfect scaling properties will be essential for proving the results in this paper. However, they do at first sight seem to prove that Thomas-Fermi theory is useless for describing the periodicity of (large) atoms by implying
\begin{equation}\label{TFLimit}
\Phi_Z^{\text{TF}}(x)\longrightarrow\Bigl(\frac{5c_{\text{TF}}}{3}\Bigr)^3\frac{9}{\uppi^2\abs{x}^4}=:\Phi_\infty^{\text{TF}}(x)\quad\text{and similarly}\quad\rho_Z^{\text{TF}}\longrightarrow\rho_\infty^{\text{TF}}
\end{equation}
pointwise on $\mathbb{R}^3\backslash\{0\}$ as $Z\to\infty$. Crucially, the $Z$ can converge towards $\infty$ in any possible way. The latter convergence can be interpreted as the exact opposite of periodicity of large atoms in this model: It says that the distribution of the electrons in the atom looks similar for large $Z$ regardless of how these are chosen. To detect a periodicity we need thus to consider a slightly more advanced model.

Let us now define a mean-field model based on Thomas-Fermi theory. The asymptotics of $\Phi_Z^{\text{TF}}$ near the origin and infinity together with its continuity yield straightforwardly $\Phi_Z^{\text{TF}}\in L^2$. Consequently, the Schrödinger operator
\[
H_Z^{\text{TF}}:=-\Delta-\Phi_Z^{\text{TF}}
\]
acting on\footnote{A more physical choice of Hilbert space would be $L^2(\mathbb{R}^3;\mathbb{C}^2)$ including spin degrees of freedom. Observe, however, that $H_Z^{\text{TF}}$ acting on this Hilbert space is unitarily equivalent to $H_Z^{\text{TF}}\oplus H_Z^{\text{TF}}$ acting on $L^2(\mathbb{R}^3)\oplus L^2(\mathbb{R}^3)$. Thus, nothing qualitative is gained by considering the larger Hilbert space.} $L^2(\mathbb{R}^3)$ is essentially self-adjoint on $C_0^\infty(\mathbb{R}^{3})$ by Kato's theorem. Its self-adjoint closure on this space is the \emph{\textbf{Thomas-Fermi mean-field model for the atom}}. We present below our findings concerning the convergence properties of these operators as $Z\to\infty$. The results use in their formulation the concepts of strong resolvent and norm resolvent convergence of self-adjoint operators. By definition, a sequence of self-adjoint operators $\{A_n\}_{n=1}^\infty$ on some fixed Hilbert space $\mathcal{H}$ converges towards another such operator $A$ in the strong resolvent sense or norm resolvent sense if the (bounded) resolvent operators $(A_n+i)^{-1}$ converge towards the corresponding $(A+i)^{-1}$ in the strong or norm sense respectively. This generalizes strong and norm convergence of bounded operators. For more details on these types of convergence, see \cite{ReedSimon1} VIII.7.

\subsection{Main result on the Thomas-Fermi mean-field model}
Unlike the situation in (\ref{TFLimit}) there is no general convergence of $H_Z^{\text{TF}}$ as $Z\to\infty$. Rather, one must choose particular sequences $\{Z_n\}_{n=1}^\infty$ of atomic numbers and consider the corresponding sequences of atoms $\{H_{Z_n}^{\text{TF}}\}_{n=1}^\infty$ in order to have strong resolvent convergence of the operators as $n\to\infty$. Concretely, we introduce the "classical constant"
\[
D_{\text{cl}}:=\frac{1}{4\uppi^2}\int\frac{\Phi_1^{\text{TF}}(x)^{1/2}}{\abs{x}^2}\,dx
\]
and obtain the following result (Theorem \ref{MainResultTF}): Suppose $Z_n\to\infty$. Then $\{H_{Z_n}^{\text{TF}}\}_{n=1}^\infty$ is converging in the strong resolvent sense if and only if 
\begin{equation}
    Z_n^{1/3}D_{\text{cl}}\to\tau\quad \text{modulo } 1,
\end{equation} for some number $\tau$ which can be taken to be in $[0,1)$. In the affirmative case,
\begin{equation}\label{PeriodicConv}
H_{Z_n}^{\text{TF}}\longrightarrow H_{\infty,\tau}^{\text{TF}}
\end{equation}
where $\{H_{\infty,\tau}^{\text{TF}}\}_{\tau\in[0,1)}$ is a parametrized family of self-adjoint extensions of the operator $-\Delta-\Phi_\infty^{\text{TF}}$ defined on $C_0^\infty(\mathbb{R}^3\backslash\{0\})$. When rewriting the convergence condition imposed on $D_{\text{cl}}Z_n^{1/3}$ we see that it is satisfied for sequences similar to $Z_\ell(n)$ in (\ref{ZellFormula}) but with the coefficient $D_{\text{cl}}^{-3}$ instead of $1/6$ on the leading $n^3$-term (cf. the remark following Definition \ref{TheDefinition} below). In this sense we recover the periodicity lost in the Thomas-Fermi density functional theory -- while our model, however, does not satisfy Madelung's rule asymptotically.

We show also, by providing a counterexample, that in the result in Theorem \ref{MainResultTF} described above one cannot generally replace "in the strong resolvent sense" with "in the norm resolvent sense". That is, there exists a sequence $\{Z_n\}_{n=1}^\infty$ so that $Z_n^{1/3}D_{\text{cl}}\to\tau$ modulo $1$ and $Z_n\to\infty$ while $\{H_{Z_n}^{\text{TF}}\}_{n=1}^\infty$ is not converging in the norm resolvent sense.

The $H_{\infty,\tau}^{\text{TF}}$'s are distinct for different $\tau$'s, and they are naturally interpreted as the infinitely many different "kinds" of infinite atoms in the Thomas-Fermi mean-field model -- corresponding to the groups in the periodic table of the (finite) atoms. Changing the parametrization to $t=(\cos2\uppi\tau,\sin2\uppi\tau)$ one obtains a continuous parametrization of the operators by the unit circle $S^1$ (see Corollary \ref{ContinuousPara}), thus recovering a periodicity aspect even for infinite atoms in this model. We note that the possible limiting operators $\{H_{\infty,\tau}^{\text{TF}}\}_{\tau\in[0,1)}$ in (\ref{PeriodicConv}) is by no means an exhaustive list of possible self-adjoint realizations of $-\Delta-\Phi_\infty^{\text{TF}}$. Even among realizations that commute with the orthogonal projections onto all angular momentum eigenspaces of the Laplace operator there is a family of distinct realizations parametrized by $S^1\times\mathbb{N}_0$. In this sense the nature of the finite Thomas-Fermi atoms singles out a specific 1-parameter family of "infinite Thomas-Fermi atoms" in a non-trivial way.

In Section \ref{sec2} we present the results described above in a more general set-up which in particular highlights the crucial properties of the Thomas-Fermi potential: Its asymptotic behaviour at the origin and infinity, and its perfect scaling in $Z$. The proofs of these results are found in Section \ref{sec3}. Finally, Section \ref{sec4} presents the promised example of a sequence of finite atoms in the Thomas-Fermi mean-field model which converges in the strong resolvent sense but not in the norm resolvent sense.
\section{Main results in a general setting}\label{sec2}
We start now the description of our main result in a form which is slightly more general than the one presented in the introduction. As a first step, we introduce a class of potentials $\Phi$ that we allow to play the role corresponding to the Thomas-Fermi potential $\Phi^\text{TF}_1$ above.
\begin{assumptions}\label{Assumptions}
We consider a radially symmetric\footnote{We will generally use the same notation for the $3$-dimensional and the $1$-dimensional radial part. It is the intent that the meaning is clear from the context throughout the material.} potential $\Phi\colon\mathbb{R}^3\to\mathbb{R}$. The assumptions on $\Phi$ (considered as a function of one variable; the radius $r$) will be
\begin{itemize}
\item[1)] $\Phi$ is strictly positive,
\item[2)] $\Phi(r)=C_0r^{\alpha}+o(r^{\alpha})$ as $r\to0$ for some $\alpha>-2$ and $C_0>0$,
\item[3)] $\Phi(r)=C_\infty r^{\beta}+o(r^{\beta})$ as $r\to\infty$ for some $\beta<-2$ and $C_\infty>0$,
\item[4)] $\Phi$ is twice continuously differentiable on $\mathbb{R}_+$,
\item[5)] $r\abs{\Phi'(r)},r^2\abs{\Phi''(r)}\lesssim r^{\alpha}$ on $(0,1)$ and $r\abs{\Phi'(r)},r^2\abs{\Phi''(r)}\lesssim r^{\beta}$ on $(1,\infty)$.
\end{itemize}
\end{assumptions}
For the remaining part of the present section we mean by $\Phi$ a potential which satisfies these assumptions. Notice that $\Phi^\text{TF}_1$ is an example of such with $\alpha=-1>-2$ and $\beta=-4<-2$ (here 5) can be verified for example by the help of 2), 3) and the differential equation in (\ref{TFEquation})). We define generally the potential \begin{equation}
    \Phi_\kappa(x):=\kappa^{-\beta}\Phi(\kappa x),
\end{equation} 
for each $\kappa>0$ -- our choice of "large parameter". Once again we can recover the situation from the introduction: In this notation the Thomas-Fermi potential $\Phi_Z^\text{TF}$ will due to its scaling properties simply be written as $(\Phi^\text{TF}_1)_{Z^{1/3}}$, i.e. $\Phi=\Phi^\text{TF}_1$ and $\kappa=Z^{1/3}$. Finally, since $\Phi_\kappa\to C_\infty\abs{x}^\beta$ as $\kappa\to\infty$ in a rather strong sense (except at the origin), we put $\Phi_\infty(x):=C_\infty\abs{x}^\beta$.

The next task is now to define the operators on $L^2(\mathbb{R}^3)$ which will model finite and infinite atoms respectively. As it is explained in the introduction, these should act as $-\Delta-\Phi_\kappa$ and $-\Delta-\Phi_\infty$ respectively, but, as it is often the case, determining their domains of self-adjointness is a more delicate matter -- especially in the infinite case. Even though we have other methods for the finite case, we present now a general construction through an angular momentum decomposition and apply this in all cases. The reason for doing so is twofold: Firstly, we really do need the construction for the infinite case, so we have to cover it anyway; secondly, the similar structure of operators describing finite and infinite atoms is crucial for the proofs of the main results below.

For the general discussion we consider an abstract radially symmetric potential $V$ which we assume is continuous. The first key idea in the construction is to separate the radial and angular variables using the standard identification $L^2(\mathbb{R}^3)\simeq L^2(\mathbb{R}_+)\otimes L^2(S^2)$ via the map $U\psi(r,\omega):=r\psi(r\omega)$ (which is a multiple of a unitary map). Notice then that by writing the Laplace operator in polar coordinates, $\Delta=r^{-1}\partial_r^2 r+r^{-2}\Delta_{S^2}$ with $\Delta_{S^2}$ the Laplace-Beltrami operator on $S^2$, one gets
\begin{equation}\label{SphericalLaplace}
U(-\Delta-V)U^{-1}(\phi\otimes\psi)=\bigl(-\frac{d^2}{dr^2}-V\bigr)\phi\otimes\psi+r^{-2}\phi\otimes(-\Delta_{S^2}\psi)
\end{equation}
for, say, $\phi\in C_0^\infty(\mathbb{R}_+)$ and $\psi\in C^\infty(S^2)$. Consequently, it is a very natural next step to further decompose the Hilbert space by using the spherical harmonics $Y_\ell^m\in \psi\in C^\infty(S^2)$, $\ell\in\mathbb{N}_0$, $m=-\ell,\dots,\ell$, which satisfy $-\Delta_{S^2}Y_\ell^m=\ell(\ell+1)Y_\ell^m$ and can be chosen so that they constitute an orthonormal basis of $L^2(S^2)$. For $\phi\in C_0^\infty(\mathbb{R}_+)$ and $\psi\in\operatorname{span}_{m=-\ell,\dots,\ell}Y_\ell^m$ we see that (\ref{SphericalLaplace}) reads $U(-\Delta-V)U^{-1}(\phi\otimes\psi)=L_\ell\phi\otimes\psi$ with
\begin{equation}\label{1D-Operator-Expression}
L_\ell=-\frac{d^2}{dr^2}+\frac{\ell(\ell+1)}{r^2}-V,
\end{equation}
and it is using this structure we define our operators rigorously below in Definition \ref{AtomDefinition}. Before doing so, we do, however, also need to handle the problem of obtaining from the expression (\ref{1D-Operator-Expression}) a self-adjoint operator on $L^2(\mathbb{R}_+)$.

To this end we define $H_{\kappa,\ell,\text{min}}$ and $H_{\infty,\ell,\text{min}}$ to be the closures of the symmetric operators acting as (\ref{1D-Operator-Expression}) on $C_0^\infty(\mathbb{R}_+)$ with $V=\Phi_\kappa$ and $V=\Phi_\infty$ respectively. By von Neumann's criterion all of these have self-adjoint extensions since they commute with complex conjugation. Moreover, the self-adjoint extensions are well understood by Weyl's limit point/limit circle criterion and the theory of generalized boundary conditions in 1-dimensional space. We now describe the results we need from these methods and refer for the details to for example the appendix to X.1 in \cite{ReedSimon2} and Appendix A in \cite{JanD1}.

A potential $W\colon\mathbb{R}_+\to\mathbb{R}$ is said to be in the \emph{\textbf{limit circle case}} at the origin and/or at infinity if all solutions to the equation $f''=Wf$ are in $L^2((0,1))$ and/or $L^2((1,\infty))$ respectively. Otherwise, it is said to be in the \emph{\textbf{limit point case}} at the origin/at infinity. It is a fundamental result by Weyl that the operator $-d^2/dr^2+W$ is essentially self-adjoint on $C_0^\infty(\mathbb{R}_+)$ if and only if $W$ is in the limit point case at both the origin and infinity. If this is not the case, then the self-adjoint extensions are defined by restricting the adjoint of the operator on $C_0^\infty(\mathbb{R}_+)$ to a smaller domain by putting (generalized) boundary conditions at the places where the potential is in the limit circle case (i.e. at the origin and/or at infinity). In our situations with $W=\ell(\ell+1)/r^2-\Phi_\kappa$ and $W=\ell(\ell+1)/r^2-\Phi_\infty$ basic estimates using 2) and 3) in Assumptions \ref{Assumptions} show that:
\begin{itemize}
\item All potentials are in the limit point case at infinity (cf. \cite{ReedSimon2}, Theorem X.8).
\item For $\ell=1,2,\dots$, the potentials $\ell(\ell+1)/r^2-\Phi_\kappa$ are in the limit point case at the origin (cf. \cite{ReedSimon2}, Theorem X.10).
\end{itemize}
Thus, letting $H_{\kappa,\ell}:=H_{\kappa,\ell,\text{min}}$ for $\ell=1,2,\dots$, these are themselves the desired self-adjoint extensions. We have further:
\begin{itemize}
\item The potentials $\ell(\ell+1)/r^2-\Phi_\infty$ for all $\ell$ are in the limit circle case at the origin (since they are decreasing as $r\to0$ from the right). Also, $-\Phi_\kappa$ is in the limit circle case\footnote{This is not entirely straightforward to realize. One way to do so is to check the definition of being in the limit circle case directly via refined knowledge about the Cauchy problem mentioned below combined with the fact that $r\mapsto r\,\Phi_\kappa(r)$ is integrable near the origin.}.
\end{itemize}
Consequently, we need in the remaining cases the notion of generalized boundary conditions. This gives, for a $W$ that is limit point at infinity and limit circle at the origin, the following characterization of all self-adjoint extensions $L_f$ of the closure of $-d^2/dr^2+W$ on $C_0^\infty(\mathbb{R}_+)$ (briefly denoted $L_\text{min}$): Take as domain the set $D(L_f):=D(L_\text{min})\oplus\mathbb{C}\xi f$ where $f$ is a real-valued solution to $f''=Wf$ and $\xi$ is a smooth localizing function which is, say, $1$ on $(0,1)$ and $0$ on $(2,\infty)$\footnote{For our entire presentation we mean by $\xi$ a fixed choice of such localizing function.}, and let $L_f$ act as $-d^2/dr^2+W$ in the distributional sense. That is, self-adjoint extensions of $H_{\kappa,0,\text{min}}$ and the $H_{\infty,\ell,\text{min}}$'s are in one-to-one correspondence with real-valued solutions to $f''=-\Phi_\kappa f$ and
\begin{equation}\label{InfiniteExtEquation}
f''(r)=\bigl[\frac{\ell(\ell+1)}{r^2}-C_\infty r^\beta\bigr]\,f(r)
\end{equation}
respectively. For extending $H_{\kappa,0,\text{min}}$ we use the fact that $r\mapsto r\,\Phi_\kappa(r)$ is in $L^1((0,1))$ so that there is a unique solution $f\in C^1([0,\infty))$ to $f''=-\Phi_\kappa f$ satisfying $f(0)=0$ and $f'(0)=1$. See \cite{JanD2} Proposition 2.5 for a proof (this problem is a special case of "the Cauchy problem"). We define $H_{\kappa,0}$ to be the self-adjoint extension of $H_{\kappa,0,\text{min}}$ obtained by choosing $D(H_{\kappa,0})=D(H_{\kappa,0,\text{min}})\oplus\mathbb{C}\xi f$ with the distinguished $f$ just described (which happens to be real-valued). In the light of Proposition \ref{PropertiesOfOperators} (a) below this turns out to be a very natural choice, agreeing with the definition of $H_Z^\text{TF}$ above. The equation (\ref{InfiniteExtEquation}) can be solved explicitly with the space of solutions spanned by the real-valued\footnote{Recall that $2+\beta<0$.} functions
\[
F_{\beta,C_\infty,\ell}(r):=\sqrt{r}\cdot J_{\frac{2\ell+1}{2+\beta}}\Bigl(\frac{-2C_\infty^{1/2}}{2+\beta}r^{\frac{2+\beta}{2}}\Bigr)\quad\text{and}\quad G_{\beta,C_\infty,\ell}(r):=\sqrt{r}\cdot Y_{\frac{2\ell+1}{2+\beta}}\Bigl(\frac{-2C_\infty^{1/2}}{2+\beta}r^{\frac{2+\beta}{2}}\Bigr)
\]
where $J$ and $Y$ are Bessel functions of the first and second kind respectively. A general reference to the properties of Bessel functions we need, including a treatment of some differential equations very similar to the ones just discussed, is \cite{AS}. An examination of the solution space described above shows that no distinguished solutions (near the origin) exist, and thus the best we can do is to come up with a convenient parametrization. Our choice is the following: For each $\ell\in\mathbb{N}_0$ and $\theta_\ell\in[0,\uppi)$ we define $H_{\infty,\ell,\theta_\ell}$ to be the self-adjoint extension of $H_{\infty,\ell,\text{min}}$ obtained by choosing
\begin{equation}\label{eq:hinftydef}
D(H_{\infty,\ell,\theta_\ell})=D(H_{\infty,\ell,\text{min}})\oplus\mathbb{C}\xi(\cos\theta_\ell F_{\beta,C_\infty,\ell}+\sin\theta_\ell G_{\beta,C_\infty,\ell}).
\end{equation}
This finishes the discussion concerning the needed self-adjoint extensions of the 1-dimensional operators. We are now in a position to define the Schrödinger operators which describe finite and infinite atoms. Recall that the motivation for the definition below is (\ref{SphericalLaplace}) together with the surrounding discussion -- in particular the line just above (\ref{1D-Operator-Expression}).
\begin{definition}\label{AtomDefinition}
We define the Schrödinger operators describing finite atoms by setting $H_\kappa=U^{-1}\widetilde{H}_\kappa U$ where $\widetilde{H}_\kappa$ is the closure of the operator $\widetilde{H}_\kappa^0$ on $L^2(\mathbb{R}_+)\otimes L^2(S^2)$ given by
\[
D(\widetilde{H}_\kappa^0)=\Bigl\{\sum_{\ell=0}^{M}\sum_{m=-\ell}^{\ell}\phi_\ell^m\otimes Y_\ell^m\;\Big\vert\; M\in\mathbb{N},\quad\phi_\ell^m\in D(H_{\kappa,\ell})\Bigr\},
\]
\[
\widetilde{H}_\kappa^0\sum_{\ell=0}^{M}\sum_{m=-\ell}^{\ell}\phi_\ell^m\otimes Y_\ell^m=\sum_{\ell=0}^{M}\sum_{m=-\ell}^{\ell}H_{\kappa,\ell}\phi_\ell^m\otimes Y_\ell^m
\]
with the self-adjoint operators $H_{\kappa,\ell}$ defined as above.

Similarly, we define Schrödinger operators by, for each sequence $\{\theta_\ell\}_{\ell=0}^{\infty}\subseteq[0,\uppi)$, setting $H_{\infty,\{\theta_\ell\}_{\ell=0}^\infty}=U^{-1}\widetilde{H}_{\infty,\{\theta_\ell\}_{\ell=0}^\infty} U$ where $\widetilde{H}_{\infty,\{\theta_\ell\}_{\ell=0}^\infty}$ is the closure of the operator $\widetilde{H}_{\infty,\{\theta_\ell\}_{\ell=0}^\infty}^0$ on $L^2(\mathbb{R}_+)\otimes L^2(S^2)$ given by
\[
D\bigl(\widetilde{H}_{\infty,\{\theta_\ell\}_{\ell=0}^\infty}^0\bigr)=\Bigl\{\sum_{\ell=0}^{M}\sum_{m=-\ell}^{\ell}\phi_\ell^m\otimes Y_\ell^m\;\Big\vert\; M\in\mathbb{N},\quad\phi_\ell^m\in D(H_{\infty,\ell,\theta_\ell})\Bigr\},
\]
\[
\widetilde{H}_{\infty,\{\theta_\ell\}_{\ell=0}^\infty}^0\sum_{\ell=0}^{M}\sum_{m=-\ell}^{\ell}\phi_\ell^m\otimes Y_\ell^m=\sum_{\ell=0}^{M}\sum_{m=-\ell}^{\ell}H_{\infty,\ell,\theta_\ell}\phi_\ell^m\otimes Y_\ell^m
\]
with the self-adjoint operators $H_{\infty,\ell,\theta_\ell}$ defined as above.
\end{definition}
The operators just described have the following convenient properties:
\begin{pro}\label{PropertiesOfOperators}\hfill
\begin{itemize}
\item[a)] For each $\kappa>0$ the operator $H_\kappa$ in Definition \ref{AtomDefinition} is self-adjoint and coincides with the Friedrichs' extension of $-\Delta-\Phi_\kappa$ on $C_0^\infty(\mathbb{R}^3\backslash\{0\})$. If moreover $\alpha>-3/2$ this in turn  coincides with the closure of $-\Delta-\Phi_\kappa$ on $C_0^\infty(\mathbb{R}^3)$.
\item[b)] For each sequence $\{\theta_\ell\}_{\ell=0}^{\infty}\subseteq[0,\uppi)$ the operator $H_{\infty,\{\theta_\ell\}_{\ell=0}^\infty}$ in Definition \ref{AtomDefinition} is a self-adjoint extension of $-\Delta-\Phi_\infty$ on $C_0^\infty(\mathbb{R}^3\backslash\{0\})$.  Additionally, $H_{\infty,\{\theta_\ell\}_{\ell=0}^\infty}\neq H_{\infty,\{\theta'_\ell\}_{\ell=0}^\infty}$ whenever $\{\theta_\ell\}_{\ell=0}^\infty\neq\{\theta'_\ell\}_{\ell=0}^\infty$.
\end{itemize}
\end{pro}
The proof of this is somewhat technical but straightforward. The details can be seen in Appendix B of \cite{Thesis}.

Now, having introduced rigorously the framework for our general model, the main results can be formulated. Recall that strong resolvent convergence of the $H_{\kappa_n}$'s means strong convergence of the resolvent operators $(H_{\kappa_n}+i)^{-1}$. We present firstly the general result and then specialize to the case of the Thomas-Fermi mean-field model discussed in the introduction above.
\begin{thm}\label{MainResult}
Consider a sequence $\{\kappa_n\}_{n=1}^\infty$ of positive real numbers such that $\kappa_n\to\infty$ as $n\to\infty$. The corresponding sequence of operators $\{H_{\kappa_n}\}_{n=1}^\infty$ is convergent in the strong resolvent sense if and only if
\begin{equation}\label{MainResultEquation}
\frac{1}{\uppi}\int_{0}^{\infty}\Phi_{\kappa_n}^{1/2}\,dr=\frac{\kappa_n^{-\frac{\beta}{2}-1}}{\uppi}\int_{0}^{\infty}\Phi_1^{1/2}\,dr\longrightarrow\tau\quad\textup{(mod $\;1$)}
\end{equation}
as $n\to\infty$ for some real number $\tau$. In the affirmative case the limiting operator is $H_{\infty,\{\theta_\ell\}_{\ell=0}^\infty}$ from Definition \ref{AtomDefinition} with
\[
\frac{\theta_\ell}{\uppi}=\tau-\frac{2\ell+1}{4+2\alpha}-\frac{2\ell+1}{4+2\beta}-\frac{1}{2}\quad\textup{(mod $\;1$)}.
\]
\end{thm}
\begin{thm}\label{MainResultTF}
Consider a sequence $\{Z_n\}_{n=1}^\infty$ of positive real numbers such that $Z_n\to\infty$ as $n\to\infty$. The corresponding sequence of operators $\{H_{Z_n}^\textup{TF}\}_{n=1}^\infty$ is convergent in the strong resolvent sense if and only if
\begin{equation}\label{MainResultEquationTF}
D_{\textup{cl}}Z_n^{1/3}=\frac{Z_n^{1/3}}{\uppi}\int_{0}^{\infty}(\Phi_1^\textup{TF})^{1/2}\,dr\longrightarrow\tau\quad\textup{(mod $\;1$)}
\end{equation}
as $n\to\infty$ for some real number $\tau$. In the affirmative case the limiting operator is $H_{\infty,\{\theta_\ell\}_{\ell=0}^\infty}$ defined as in Definition \ref{AtomDefinition} with $\Phi_\infty=\Phi_\infty^\textup{TF}$ and
\[
\frac{\theta_\ell}{\uppi}=\tau+\frac{\ell}{2}+\frac{1}{4}\quad\textup{(mod $\;1$)}.
\]
In particular, this act in the $\ell^{\textup{th}}$ angular momentum sector as the self-adjoint operator
\[
H_{\infty,\ell,\theta_\ell}=-\frac{d^2}{dx^2}+\frac{\ell(\ell+1)}{\abs{x}^2}-C_\infty\abs{x}^{-4}
\]
with $C_\infty=(5c_{\textup{TF}}/3)^3\cdot(9/\uppi^2)$ and domain given by
\[
D(H_{\infty,\ell,\textup{min}})\oplus\mathbb{C}\xi\bigl(\sin\bigl(\tau\uppi+\frac{\ell\uppi}{2}+\frac{\uppi}{4}\bigr) j_\ell(C_\infty^{1/2}r^{-1})-\cos\bigl(\tau\uppi+\frac{\ell\uppi}{2}+\frac{\uppi}{4}\bigr) y_\ell(C_\infty^{1/2}r^{-1})\bigr)
\]
where $j_\ell$ and $y_\ell$ are the spherical Bessel functions of the first and second kind respectively. With our choice of units, $C_\infty=81\uppi^2$.
\end{thm}
\begin{definition}\label{TheDefinition}
We call the Schrödinger operators $H_{\infty,\{\theta_\ell\}_{\ell=0}^{\infty}}$ that appear as limits of finite atoms in Theorem \ref{MainResult} \textit{\textbf{infinite atoms}}. Similarly, we define an \textit{\textbf{infinite Thomas-Fermi mean-field atom}} to be one of the limiting operators in Theorem \ref{MainResultTF}.
\end{definition}
\textbf{Remark.} In Theorem \ref{MainResultTF} it seems a natural question to ask whether all infinite Thomas-Fermi mean-field atoms arise as strong resolvent limits of finite atoms with integer atomic numbers $Z_n$. This is indeed the case, and one can for example choose
\begin{equation}\label{IntegerCharge}
Z_n=\bigl\lfloor D_{\text{cl}}^{-3}(n+\tau)^3\bigr\rfloor
\end{equation}
to obtain the convergence (\ref{MainResultEquationTF}). More generally, taking these $Z_n$'s and adding to them a term behaving like $Cn^2+o_{n\to\infty}(n^2)$ for large $n$ results in new sequence also satisfying (\ref{MainResultEquationTF}) with $\tau+(CD_{\text{cl}}^3)/3$ instead of $\tau$.
\bigskip

\noindent Our method to prove Theorem \ref{MainResult} relies on approximating the zero-energy solutions $f$ to Schr\"odinger equations
$(-\Delta-\lambda^2\Phi)f=0$ for 
large $\lambda$ with $\Phi$ satisfying Assumptions \ref{Assumptions}. To do this we use a variant of the JWKB approximation. As usual we need to consider approximations to the solution in different regions. Because of the positivity of $\Phi$ and its behavior near the origin we, however, cannot use the standard way of matching in terms of Airy functions. Instead we match the oscillations of solutions in the overlap regions. As a byproduct we get the following general result (recall the splitting of $-\Delta-\lambda^2\Phi$ into the angular momentum operators given in (\ref{1D-Operator-Expression})).
\begin{thm}\label{JWKB-general}
Assume that $\Phi$ is a potential satisfying 1), 2), 4) and the first half of 5) in Assumptions \ref{Assumptions}, and let $\ell\in\mathbb{N}_0$ be given. Then there are unique solutions $w_{\lambda,\ell}$ to the problems\footnote{For uniqueness it is enough to have only the first condition as $x\to 0$.} 
\begin{equation}\label{StandardWKBEq}
\begin{dcases*}
w_{\lambda,\ell}''(x)=\bigl[\frac{\ell(\ell+1)}{x^2}-\lambda^2\Phi(x)\bigr]w_{\lambda,\ell}(x)
\\
x^{-\ell-1}w_{\lambda,\ell}(x)\to1\qquad\text{as }x\to0
\\
x^{-\ell}w'_{\lambda,\ell}(x)\to\ell+1\qquad\text{as }x\to0,
\end{dcases*}
\end{equation}
and these can be written in the form
\[
w_{\lambda,\ell}(x)=a_{\lambda,\ell}\Phi(x)^{-1/4}\Bigl(\cos\Bigl(\lambda\int_{0}^{x}\Bigl[\Phi(y)-\frac{\bigl(\ell+\frac{1}{2}\bigr)^2}{\lambda^2y^2}\Bigr]_+^{1/2}\,dy-\frac{\uppi}{4}\Bigr)+o_{\lambda\to\infty}(1)\Bigr)
\]
on any compact subinterval of $(0,\infty)$ with $o_{\lambda\to\infty}(1)$ uniform on this interval. Here $a_{\lambda,\ell}$ are constants.
\end{thm}
It is worth noticing that while  we assume 2) of Assumptions~\ref{Assumptions}  in Theorem~\ref{JWKB-general} the $\alpha$ does not appear in the statement. It is therefore natural to ask whether this assumption can be relaxed. 

Another point to notice is that the angular momentum $\ell$ appears in the approximation above in the expression $(\ell+\frac12)^2$ rather than the $\ell(\ell+1)$ that appears in (\ref{StandardWKBEq}). That this is a better approximation was known for long in the physics literature and Langer \cite{Langer} gave a heuristic explanation.  We here give a rigorous analysis relying on the Langer transformation mapping $\mathbb{R}_+$ to $\mathbb{R}$ introduced in \cite{Langer}.  It is remarkable that the universal $\frac\pi4$ in the approximation above is the same as what appears in the general JWKB approximation, although the derivation here is very different. We remark that an alternative rigorous approach reaching some of the same conclusions, but for a much smaller class of potentials not including the Thomas-Fermi potential, is discussed in \cite{ExactWKB}.

Theorem~\ref{JWKB-general} is a slightly weaker version of Theorem~\ref{JWKBLanger} below. In Theorem~\ref{JWKBLanger} the interval on which uniform approximation holds is allowed a sufficiently slow growth as function of $\lambda$ both towards 0 and $\infty$. Theorem~\ref{JWKB-general} above is, indeed,  also correct if the interval is allowed to approach zero in the manner described in Theorem~\ref{JWKBLanger}. Allowing also a growth toward infinity would require 3) and the second half of 5) in Assumptions~\ref{Assumptions}. 
\section{Proofs}\label{sec3}
\subsection{First reductions}\label{subsec3.1}
To reduce the problem of proving the "if"-part of Theorem \ref{MainResult} to a more concrete convergence problem, we introduce abstractly the notion of the \emph{\textbf{strong limit of the graphs}}, $\Gamma(A_n)$, of a sequence of operators $\{A_n\}_{n=1}^\infty$ on a fixed Hilbert space $\mathcal{H}$. That is, we let $\text{str.lim }\Gamma(A_n)$ be the set of $(\phi,\psi)\in\mathcal{H}\times\mathcal{H}$ satisfying that there exist $\phi_n\in D(A_n)$ so that $\phi_n\to\phi$ and $A_n\phi_n\to\psi$ in the Hilbert space as $n\to\infty$. This concept is closely related to strong graph convergence of operators which is discussed in for example \cite{ReedSimon1} VIII.7. A diagonal argument shows that strong limits of graphs are closed subspaces of $\mathcal{H}\times\mathcal{H}$ -- for the details of this and further results on strong limits of subspaces in general and of graphs in particular, we refer the reader to \cite{StrongGraphConvergence}. We do, however, in Lemma \ref{GraphLemma1} and Proposition \ref{GraphLemma2} provide proofs of the properties of these limits that are essential to our proof of Theorem \ref{MainResult}. Firstly we have:
\begin{lemma}\label{GraphLemma1}
Consider sequences $\{\kappa_n\}_{n=1}^\infty\subseteq\mathbb{R}_+$ and $\{\theta_\ell\}_{\ell=0}^\infty\subseteq[0,\uppi)$. If
\begin{equation}\label{GraphInclusion}
\Gamma\bigl(\widetilde{H}_{\infty,\{\theta_\ell\}_{\ell=0}^\infty}^0\bigr)\subseteq \textup{str.lim }\Gamma(\widetilde{H}_{\kappa_n}^0)
\end{equation}
then $H_{\kappa_n}\to H_{\infty,\{\theta_\ell\}_{\ell=0}^\infty}$ in the strong resolvent sense as $n\to\infty$.
\end{lemma}
\proof
We observe that strong resolvent convergence of $H_{\kappa_n}$ towards $H_{\infty,\{\theta_\ell\}_{\ell=0}^\infty}$ is clearly equivalent to that of $\widetilde{H}_{\kappa_n}$ towards $\widetilde{H}_{\infty,\{\theta_\ell\}_{\ell=0}^\infty}$. Consider for the sake of proving the latter any function $\psi\in L^2(\mathbb{R}_+)\otimes L^2(S^2)=R(\widetilde{H}_{\infty,\{\theta_\ell\}_{\ell=0}^\infty}+i)$ and write this as $\psi=(\widetilde{H}_{\infty,\{\theta_\ell\}_{\ell=0}^\infty}+i)\phi$ for some $\phi\in D(\widetilde{H}_{\infty,\{\theta_\ell\}_{\ell=0}^\infty})$. Notice now that with the assumption (\ref{GraphInclusion}) we have
\[
\Gamma\bigl(\widetilde{H}_{\infty,\{\theta_\ell\}_{\ell=0}^\infty}\bigr)=\overline{\Gamma\bigl(\widetilde{H}_{\infty,\{\theta_\ell\}_{\ell=0}^\infty}^0\bigr)}\subseteq\textup{str.lim }\Gamma(\widetilde{H}_{\kappa_n}^0)\subseteq\textup{str.lim }\Gamma(\widetilde{H}_{\kappa_n}),
\]
which means that there exist some $\phi_n\in D(\widetilde{H}_{\kappa_n})$ satisfying both $\phi_n\to\phi$ and $\widetilde{H}_{\kappa_n}\phi_n\to \widetilde{H}_{\infty,\{\theta_\ell\}_{\ell=0}^\infty}\phi$. Consequently,
\begin{align*}
\bigl[(\widetilde{H}_{\kappa_n}+i)^{-1}&-(\widetilde{H}_{\infty,\{\theta_\ell\}_{\ell=0}^\infty}+i)^{-1}\bigr]\,\psi
\\
&=(\widetilde{H}_{\kappa_n}+i)^{-1}\bigl[(\widetilde{H}_{\infty,\{\theta_\ell\}_{\ell=0}^\infty}+i)\phi-(\widetilde{H}_{\kappa_n}+i)\phi_n\bigr]-\phi+\phi_n\longrightarrow0,
\end{align*}
where we used the fact that $\norm{(\widetilde{H}_{\kappa_n}+i)^{-1}}\leq1$ for all $n$. This finishes the proof.\qed
We have also the following straightforward reduction of the proof:
\begin{lemma}\label{EasyImplication}
In Theorem \ref{MainResult}, the "only if"-part follows from all the remaining assertions.
\end{lemma}
\proof Suppose the remaining assertions of Theorem \ref{MainResult} hold true and consider a sequence $\{\kappa_n\}_{n=1}^\infty$ of positive real numbers so that $\kappa_n\to\infty$ but the integrals
\[
\frac{1}{\uppi}\int_{0}^{\infty}\Phi_{\kappa_n}^{1/2}\,dr=:K_n
\]
is not convergent modulo $1$. Since the latter (non-)convergence takes place in a compact space, it must be the case that $\{K_n\}_{n=1}^\infty$ has at least two accumulation points $\tau\neq\tau'$ in this space, i.e. modulo $1$. Now choosing subsequences along which $\{K_n\}_{n=1}^\infty$ converges towards $\tau$ and $\tau'$ respectively, the remaining assertions of the theorem tells us that along these subsequences the corresponding Schrödinger operators converge in the strong resolvent sense towards $H_{\infty,\{\theta_\ell\}_{\ell=0}^\infty}$ and $H_{\infty,\{\theta'_\ell\}_{\ell=0}^\infty}$ respectively where $\{\theta_\ell\}_{\ell=0}^\infty\neq\{\theta'_\ell\}_{\ell=0}^\infty$. But by the last part of Proposition \ref{PropertiesOfOperators} (b) these operators are unequal, and hence this implies that $\{H_{\kappa_n}\}_{n=1}^\infty$ cannot converge towards any single operator in the strong resolvent sense.
\qed
We see from Lemmas~\ref{GraphLemma1} and \ref{EasyImplication} that in order to prove Theorem \ref{MainResult} it suffices to verify \eqref{GraphInclusion} for sequences $\kappa_n$ satisfying the conditions in Theorem~\ref{MainResult}. We now reduce this to proving more concrete convergence properties. 
\begin{pro}\label{GraphLemma2}
Given $\{\theta_\ell\}_{\ell=0}^\infty\subseteq[0,\uppi)$ and a sequence $\kappa_n\to\infty$ as $n\to\infty$. Then (\ref{GraphInclusion}) holds if there exist functions $\phi_{n,\ell}\in D(H_{\kappa_n,\ell})$ such that
\begin{itemize}
\item $\phi_{n,\ell}\longrightarrow \xi(\cos\theta_\ell F_{\beta,C_\infty,\ell}+\sin\theta_\ell G_{\beta,C_\infty,\ell})$
\item $H_{\kappa_n,\ell} \phi_{n,\ell} \longrightarrow H_{\infty,\ell,\theta_\ell}(\xi(\cos\theta_\ell F_{\beta,C_\infty,\ell}+\sin\theta_\ell G_{\beta,C_\infty,\ell}))$
\end{itemize}
in $L^2(\mathbb{R}_+)$ as $n\to\infty$ for all $\ell\in\mathbb{N}_0$. Here $F_{\beta,C_\infty,\ell}$ and $G_{\beta,C_\infty,\ell}$ are the solutions to (\ref{InfiniteExtEquation}) defined in Section \ref{sec2}.
\end{pro}
\proof
We observe first that proving $\Gamma(H_{\infty,\ell,\theta_\ell})\subseteq\text{str.lim }\Gamma(H_{\kappa_n,\ell})$ for each $\ell$ implies (\ref{GraphInclusion}). Indeed,  considering  an arbitrary element
\[
\sigma:=\sum_{\ell=0}^{M}\sum_{m=-\ell}^{\ell}\phi_\ell^m\otimes Y_\ell^m\in D\bigl(\widetilde{H}_{\infty,\{\theta_\ell\}_{\ell=0}^\infty}^0\bigr)
\]
we may take $\psi_{\ell,n}^m\in D(H_{\kappa_n,\ell})$ so that $\psi_{\ell,n}^m\to\phi_\ell^m$ and $H_{\kappa_n,\ell}\psi_{\ell,n}^m\to H_{\infty,\ell,\theta_\ell}\phi_\ell^m$ as $n\to\infty$. It is then easy to check that 
\[
\sigma_n:=\sum_{\ell=0}^{M}\sum_{m=-\ell}^{\ell}\psi_{\ell,n}^m\otimes Y_\ell^m\in D(\widetilde{H}_{\kappa_n}^0),
\]
 satisfies $\sigma_n\to \sigma$ and $\widetilde{H}_{\kappa_n}^0\sigma_n\to\widetilde{H}_{\infty,\{\theta_\ell\}_{\ell=0}^\infty}^0\sigma$ thus implying (\ref{GraphInclusion}). 

To conclude that 
$\Gamma(H_{\infty,\ell,\theta_\ell})\subseteq\text{str.lim }\Gamma(H_{\kappa_n,\ell})$ holds for all $\ell\in\mathbb{N}_0$ we use the existence of the functions $\phi_{n,\ell}$ and recall that from \eqref{eq:hinftydef}
\[
\Gamma(H_{\infty,\ell,\theta_\ell})=\Gamma(H_{\infty,\ell,\text{min}})\oplus\mathbb{C}(\lim_{n\to\infty}\phi_{n,\ell},\lim_{n\to\infty}H_{\kappa_n,\ell}\phi_{n,\ell}),
\]
where the direct sum is in $L^2(\mathbb{R}_+)\times L^2(\mathbb{R}_+)$. Here the last term is clearly a subset of $\text{str.lim }\Gamma(H_{\kappa_n,\ell})$. Moreover, we have for any $h\in C_0^\infty(\mathbb{R}_+)$ the convergence
\[
\int_{0}^{\infty}\abs{H_{\infty,\ell,\text{min}}h-H_{\kappa_n,\ell}h}^2\,dr=\int_{0}^{\infty}\abs{h}\cdot\abs{\Phi_{\kappa_n}-\Phi_\infty}^2\,dr\longrightarrow0
\]
so that also
\[
\Gamma(H_{\infty,\ell,\text{min}})=\overline{\Gamma(H_{\infty,\ell,\text{min}}\vert_{C_0^\infty(\mathbb{R}_+)})}\subseteq\overline{\text{str.lim }\Gamma(H_{\kappa_n,\ell})}=\text{str.lim }\Gamma(H_{\kappa_n,\ell})
\]
as claimed.
\qed
We observe at this point that, with the tools introduced for proving Proposition \ref{GraphLemma2}, the continuity of the parametrization of infinite atoms mentioned in the introduction follows quite easily.
\begin{cor}\label{ContinuousPara}
Denote, for each $\tau$, by $H_{\infty,\tau}$ the limiting operator described in Theorem \ref{MainResult}. If $\{\tau_n\}_{n=1}^{\infty}$ is a sequence of real numbers converging towards some $\tau_0$ modulo $1$ then the operators $H_{\infty,\tau_n}$ converge towards $H_{\infty,\tau_0}$ in the strong resolvent sense.
\end{cor}
\proof We introduce for convenience the numbers
\[
\nu_{\alpha,\beta,\ell}=\frac{(2\ell+1)\uppi}{4+2\alpha}+\frac{(2\ell+1)\uppi}{4+2\beta}+\frac{\uppi}{2},
\]
so that, with the notation from Section 2,
\[
H_{\infty,\tau_n}=H_{\infty,\{\uppi\tau_n-\nu_{\alpha,\beta,\ell}\}_{\ell=0}^\infty}\quad\text{and}\quad H_{\infty,\tau_0}=H_{\infty,\{\uppi\tau_0-\nu_{\alpha,\beta,\ell}\}_{\ell=0}^\infty}.
\]
Now, following the proofs of Lemma \ref{GraphLemma1} and Proposition \ref{GraphLemma2}, we know that it suffices to prove the inclusion $\Gamma(H_{\infty,\ell,\uppi\tau_0-\nu_{\alpha,\beta,\ell}})\subseteq\text{str.lim }\Gamma(H_{\infty,\ell,\uppi\tau_n-\nu_{\alpha,\beta,\ell}})$ for all $\ell$. Clearly, this in turn boils down to arguing that
\begin{align*}
\bigl(\xi(\cos(\uppi&\tau_0-\nu_{\alpha,\beta,\ell})F_{\beta,C_\infty,\ell}+\sin(\uppi\tau_0-\nu_{\alpha,\beta,\ell})G_{\beta,C_\infty,\ell})\;,
\\
&H_{\infty,\ell,\uppi\tau_0-\nu_{\alpha,\beta,\ell}}(\xi(\cos(\uppi\tau_0-\nu_{\alpha,\beta,\ell})F_{\beta,C_\infty,\ell}+\sin(\uppi\tau_0-\nu_{\alpha,\beta,\ell})G_{\beta,C_\infty,\ell}))\bigr)
\end{align*}
lies in $\text{str.lim }\Gamma(H_{\infty,\ell,\uppi\tau_n-\nu_{\alpha,\beta,\ell}})$. This fact is readily checked by substituting $\tau_0$ by $\tau_n$ in the above expression and letting $n\to\infty$.
\QED
\subsection{Setup and strategy}\label{Strategy}

We now begin the process of constructing the functions $\phi_{n,\ell}$ satisfying the properties in Proposition~\ref{GraphLemma2}.
In this subsection we outline the strategy and fix some notation. For the remaining part of the present section we mean by $\Phi$ a potential that satisfies the Assumptions \ref{Assumptions} and by $\xi$, $\Phi_\kappa$, $\Phi_\infty$, $H_\kappa$ and $H_{\infty,\{\theta_\ell\}_{\ell=0}^\infty}$ the quantities introduced in Section \ref{sec2}. Since we will be in a 1-dimensional setting, we change the space variable from the radial $r$ to the more standard choice of $x$. Adapting to the nature of the statement in Proposition \ref{GraphLemma2}, we also allow ourselves to treat $\ell$ simply as a constant from this point onwards. We use the notation ``$\lesssim$" to indicate ``less than up to a constant". Here the constant might depend on $\ell$ and on the potential $\Phi$, but it may not depend on $x$ or $\kappa$. In the same spirit, ``$\propto$'' always indicates ``proportional to'' as a function \emph{only} of $x$.

A natural approach to the construction of $\phi_{n,\ell}$ would be to choose them to be $c_n\xi f_{\kappa_n,\ell}$ where $c_n\ne0$ and $f_{\kappa_n,\ell}$ are solutions to the zero energy equation 
$$f_{\kappa_n,\ell}''=[\ell(\ell+1)x^{-2}-\Phi_{\kappa_n}]f_{\kappa_n,\ell}$$ satisfying appropriate boundary conditions at the origin. The second bullet point in Proposition~\ref{GraphLemma2} is then obvious close to zero.
We would now like to conclude from the assumption (\ref{MainResultEquation}) that an appropriate choice of $c_n$ gives
$$
c_nf_{\kappa_n,\ell}\longrightarrow\cos\theta_\ell F_{\beta,C_\infty,\ell}+\sin\theta_\ell G_{\beta,C_\infty,\ell} 
$$
in $L^2$ on the support of $\xi$. Unfortunately, we do not have the sufficient control on $f_{\kappa_n,\ell}$ to do so up to constant distances from the origin. To circumvent this technical problem we will replace $\Phi$ by a potential which is equal to $\Phi$ close to the origin and exactly equal to $\Phi_\infty$ further away. 

The entire analysis will be carried out after a \emph{\textbf{Langer transformation}} -- a change of variable $x\to\kappa^{-1}e^x$ -- which was first suggested in \cite{Langer} for studying the JWKB approximation in the context of the Schrödinger equation for the hydrogen atom. The fact that this problem is very similar to ours already hints that the Langer transformation might help us. Its usefulness will be more apparent below, but let us observe for the moment that if we put $\lambda={\kappa^{-(2+\beta)/2}}$ and $g_{\lambda,\ell}(x)=e^{-x/2}f_{\kappa,\ell}(\kappa^{-1}e^x)$ with $f_{\kappa,\ell}$ solving the equation above then, with $L=\ell+1/2$,
\begin{equation}\label{LangerEquation}
\begin{split}
g_{\lambda,\ell}''(x)&=\bigl[\bigl(\ell+\frac{1}{2}\bigr)^2-\kappa^{-2-\beta}e^{2x}\Phi(e^x)\bigr]g_{\lambda,\ell}(x)
\\
&=\lambda^2\bigl[L^2\lambda^{-2}-e^{2x}\Phi(e^x)\bigr]g_{\lambda,\ell}(x)=:-\lambda^2V_{\lambda,\ell}(x)g_{\lambda,\ell}(x)
\end{split}
\end{equation}
on $\mathbb{R}$.

It is in this equation that we for the technical reasons described above need to replace $\Phi$ by the potential 
\[
\Psi_\lambda:=\zeta_\lambda\Phi+(1-\zeta_\lambda)\Phi_{\infty}.
\]
The cut-off function $\zeta_\lambda$ depends on a function $\eta: {\mathbb R}_+\to{\mathbb R}_+$ which tends to infinity at infinity. We will ultimately choose this convergence to be sufficiently slow in a way so as to satisfy the requirements in Proposition \ref{LGApproximation2} and Corollaries \ref{CorrectionToSVCor} and \ref{LGApproximation2Cor} in addition to part 1 of the proof of Theorem \ref{MainResult} in Subsection \ref{ProofSubSec} below. For each $\lambda>0$ we then let $\zeta_\lambda$ be a smooth function with values in $[0,1]$ and which is $1$ on $(0,\kappa e^{-3\eta(\lambda)})$ and $0$ on $(\kappa e^{-2\eta(\lambda)},\infty)$ and additionally satisfying
\[
\sup_{\mathbb{R}_+}\abs{\zeta_\lambda'}\lesssim(\kappa e^{-2\eta(\lambda)}-\kappa e^{-3\eta(\lambda)})^{-1}\quad\text{and}\quad\sup_{\mathbb{R}_+}\abs{\zeta_\lambda''}\lesssim(\kappa e^{-2\eta(\lambda)}-\kappa e^{-3\eta(\lambda)})^{-2}.
\]
It is an easy check that such functions exist by scaling appropriately a fixed smooth function. Since $\zeta_\lambda'=\zeta_\lambda''\equiv0$ on $(\kappa e^{-2\eta(\lambda)},\infty)$, the above uniform bounds on the derivatives imply
\begin{equation}\label{ZetaBounds}
\abs{\zeta_\lambda'(x)}\lesssim\frac{1}{x}\qquad\text{and}\qquad\abs{\zeta_\lambda''(x)}\lesssim\frac{1}{x^2}.
\end{equation}
Corresponding to $\Psi_\lambda$ we put
\[
\widetilde{V}_{\lambda,\ell}(x):=e^{2x}\Psi_\lambda(e^x)-\lambda^{-2}\bigl(\ell+\frac{1}{2}\bigr)^2
\]
and consider $\widetilde{g}_{\lambda,\ell}$ to be the regular solution (see Definition~\ref{RegularSolution} below for the definition of regular solution) to 
\begin{equation}\label{eq:gtilde}
\widetilde{g}_{\lambda,\ell}''=-\lambda^2\widetilde{V}_{\lambda,\ell}\widetilde{g}_{\lambda,\ell} .
\end{equation} We then define 
$\widetilde{f}_{\kappa,\ell}$ such that $\widetilde{g}_{\lambda,\ell}(x)=e^{-x/2}\widetilde{f}_{\kappa,\ell}(\kappa^{-1}e^x)$. 
Finally, we choose $\phi_{n,\ell}(x)=c_n\xi\widetilde{f}_{\kappa_n,\ell}$ 
and prove, using assumption (\ref{MainResultEquation}), that we can choose $c_n$ such that they satisfy the properties in Proposition~\ref{GraphLemma2}. This is done by a detailed analysis of the asymptotic behaviour of the solutions $\widetilde{g}_{\lambda,\ell}$ to (\ref{eq:gtilde}).

We note that the regular solution $g_{\lambda,\ell}$ of (\ref{LangerEquation}) agrees with $\widetilde{g}_{\lambda,\ell}(x)$ for $x<\ln\kappa-3\eta(\lambda)=-\frac{2\ln\lambda}{2+\beta}-3\eta(\lambda)$. While we really need to analyze $\widetilde{g}_{\lambda,\ell}(x)$ several proofs become more natural and of independent interest for $g_{\lambda,\ell}$. We have throughout the analysis chosen to give these proofs first and formulate the results for $\widetilde{g}_{\lambda,\ell}(x)$ as corollaries.

To determine the behaviour of the solutions of (\ref{LangerEquation}) and (\ref{eq:gtilde}) we look at some regions separately: Near minus infinity we can use 2) from Assumptions \ref{Assumptions} to control solutions directly. On the next large part of the real axis, $V_{\lambda,\ell}$ and $\widetilde{V}_{\lambda,\ell}$ will be strictly positive for large $\lambda$, and 4) and 5) from Assumptions \ref{Assumptions} allow us to use the \emph{\textbf{Liouville-Green approximation}} to describe solutions here as well. Lastly, we can put an appropriate boundary condition at minus infinity and "glue together" the descriptions that we have of the solutions on these two regions. Since we know the exact form of $\widetilde{V}_{\lambda,\ell}$ (and hence of $\widetilde{g}_{\lambda,\ell}$) from some point onwards, this allows us to verify the convergences in Proposition \ref{GraphLemma2} when assuming (\ref{MainResultEquation}).
\subsection{The regular zero energy solutions}
From this point onwards we use the notation from Subsection \ref{Strategy}, i.e. $\xi$, $\Phi$, $\Phi_\kappa$, $H_{\kappa,\ell}$, $H_{\infty,\ell,\theta_\ell}$, $\lambda$, $\Psi_\lambda$, $V_{\lambda,\ell}$ and $\widetilde{V}_{\lambda,\ell}$ are as described above. As a first step in the analysis of the ``zero energy solutions'', i.e. $f_{\kappa,\ell}$, $\widetilde{f}_{\kappa,\ell}$, $g_{\lambda,\ell}$ and $\widetilde{g}_{\lambda,\ell}$ from Subsection \ref{Strategy} we prove the existence of these with certain boundary conditions. For the former this means that $\xi f_{\kappa,\ell},\xi \widetilde{f}_{\kappa,\ell}\in D(H_{\kappa,\ell})$ and for the latter that the solutions are exponentially small at minus infinity.
\begin{pro}\label{ExistenceOfSolutions}
Let $W\colon\mathbb{R}\to\mathbb{R}$ be any continuous potential satisfying
\begin{equation}\label{IntegrablePotentials}
\int_{-\infty}^{x}\abs{W(y)}\,dy=:Q(x)<\infty
\end{equation}
for all $x\in\mathbb{R}$ and let $L>0$ be any number. Then there exists a real-valued solution $g\in C^2(\mathbb{R})$ to the equation $g''=[L^2+W]g$ satisfying $e^{-Lx}g(x)\to1$ and $e^{-Lx}g'(x)\to L$ as $x\to-\infty$.
\end{pro}
\proof
The proof is constructive with the following construction of the solution $g$: Define $h_0(x)=e^{Lx}$ and then 
\[
h_i(x)=\frac{1}{L}\int_{-\infty}^{x}\sinh\bigl(L(x-y)\bigr)W(y)h_{i-1}(y)\,dy
\]
for each $i=1,2,3,\dots$. We notice that $\abs{h_i(x)}\leq e^{Lx}Q(x)^i/(L^ii!)$ for all $i\in\mathbb{N}_0$ so that this is well-defined. Indeed, this can be seen by the induction step
\begin{equation}\label{ExponentialEstimate}
\abs{h_i(x)}\leq\frac{e^{Lx}}{L}\int_{-\infty}^{x}\abs{W(y)}\frac{Q(y)^{i-1}}{L^{i-1}(i-1)!}\,dy=e^{Lx}\frac{Q(x)^i}{L^ii!},
\end{equation}
where we estimated $\sinh z\leq e^z$ for $z>0$, and this bound implies that the integral defining each $h_i$ is convergent. Also, we get from the estimate, on any interval of the form $(-\infty,x_0]$, uniform convergence of the series
\[
\sum_{i=0}^{\infty}h_i(x)
\]
towards some real-valued continuous function $g$ that satisfies $\abs{g(x)}\leq e^{Lx+Q(x)/L}$ and similarly of
\[
\sum_{i=0}^{\infty}e^{-Lx}h_i(x)
\]
towards $e^{-Lx}g(x)$. In turn this tells us that
\begin{align*}
e^{Lx}+\int_{-\infty}^{x}\frac{\sinh\bigl(L(x-y)\bigr)}{L}&W(y)g(y)\,dy
\\
&=h_0(x)+\sum_{i=0}^{\infty}\int_{-\infty}^{x}\frac{\sinh\bigl(L(x-y)\bigr)}{L}W(y)h_i(y)\,dy=g(x)
\end{align*}
so that $g$ is differentiable with
\begin{equation}\label{gPrime}
g'(x)=Le^{Lx}+\int_{-\infty}^{x}\cosh\bigl(L(x-y)\bigr)W(y)g(y)\,dy
\end{equation}
and further
\[
g''(x)=L^2e^{Lx}+W(x)g(x)+L\int_{-\infty}^{x}\sinh\bigl(L(x-y)\bigr)W(y)g(y)\,dy=[L^2+W(x)]g(x)
\]
as needed. The $C^2$-property of $g$ follows from this equation and the fact that $g$ is continuous.

For the first assertion about the limit as $x\to-\infty$ simply notice that in this limit
\[
\abs{e^{-Lx}g(x)-1}=\Bigl\vert e^{-Lx}\sum_{i=1}^{\infty}h_i(x)\Bigr\vert\leq\sum_{i=1}^{\infty}\frac{Q(x)^i}{L^ii!}=e^{Q(x)/L}-1\longrightarrow0,
\]
where we used once again the estimate (\ref{ExponentialEstimate}). For the second one observe that further
\[
\Bigl\vert e^{-Lx}\int_{-\infty}^{x}\cosh\bigl(L(x-y)\bigr)W(y)g(y)\,dy\Bigr\vert\leq e^{Q(x)/L}\int_{-\infty}^{x}\abs{W(y)}\,dy\longrightarrow 0,
\]
where we estimated $\cosh(z)\leq e^z$ for $z>0$ and $e^{-Ly}\abs{g(y)}\leq e^{Q(y)/L}$, so that (\ref{gPrime}) yields the desired conclusion.
\QED
\begin{definition}\label{RegularSolution}
For $W$ and $L$ as in Proposition \ref{ExistenceOfSolutions} we call the $g$ constructed in the proof hereof the \textbf{regular solution} of $g''=[L^2+W]g$.
\end{definition}
Notice that the asymptotics of the potentials $\Phi$ and $\Psi_\lambda$ near zero are such that the equations (\ref{LangerEquation}) and (\ref{eq:gtilde}) are of the form in Definition~\ref{RegularSolution}. It therefore is meaningful to define the functions $g_{\lambda,\ell}$ and $\widetilde g_{\lambda,\ell}$ as the regular solutions as we did above, and moreover $g_{\lambda,\ell}(x)=\widetilde g_{\lambda,\ell}(x)$ for $x<-\frac{2\ln\lambda}{2+\beta}-3\eta(\lambda)$ by construction. In Subsection \ref{Strategy} we also defined
\begin{equation}\label{fDef}
f_{\kappa,\ell}(x)=\sqrt{x}\,g_{\lambda,\ell}(\ln\kappa+\ln x),\quad\text{and}\quad
\widetilde f_{\kappa,\ell}(x)=\sqrt{x}\,\widetilde g_{\lambda,\ell}(\ln\kappa+\ln x)
\end{equation}
which have the following properties:

\begin{lemma}\label{PhiInDomain}
For each $\kappa>0$ and $\ell\in\mathbb{N}_0$ the function $f_{\kappa,\ell}$ satisfies the equation $f_{\kappa,\ell}''=[\ell(\ell+1)x^{-2}-\Phi_\kappa]f_{\kappa,\ell}$, and $\xi f_{\kappa,\ell}\in D(H_{\kappa,\ell})$ where $\xi$ is as described in Section \ref{sec2}.
\end{lemma}
\proof
The fact that $f_{\kappa,\ell}$ satisfies the equation is a straightforward calculation using the equation for $g_{\lambda,\ell}$. For the other assertion we recall that
\[
D(H_{\kappa,\ell})=D(H_{\kappa,\ell,\text{min}}^*)=\bigl\{\psi\in L^2(\mathbb{R}_+)\;\big\vert\;-\psi''+[\ell(\ell+1)x^{-2}-\Phi_\kappa]\psi\in L^2(\mathbb{R}_+)\bigr\}
\]
for $\ell=1,2,3,\dots$. It is easy to verify that $\psi=\xi f_{\kappa,\ell}$ is in this set: It is continuous, tends towards $0$ as near the origin, as is easily verified, and has support in $(0,2)$, and is thus in $L^2(\mathbb{R}_+)$. The other condition holds true since the expression that is required to be in $L^2(\mathbb{R}_+)$ is continuous in addition to being $0$ on $(0,1)$ (by the equation that $f_{\kappa,\ell}$ solves) and on $(2,\infty)$ (since $\psi\equiv0$ here).

In order to show $\xi f_{\kappa,0}\in D(H_{\kappa,0})$ it suffices by the definition of this domain to argue that $f_{\kappa,0}\in C^1([0,\infty))$ with $f_{\kappa,0}(0)=0$ (notice that since $f_{\kappa,0}$ is not identically $0$, we then cannot have $f_{\kappa,0}'(0)=0$). As mentioned above, it is easy to check that $f_{\kappa,0}\in C([0,\infty))$ with $f_{\kappa,0}(0)=0$. For the remaining part of the statement simply observe that
\begin{equation}\label{fIsRegular}
\begin{split}
f_{\kappa,0}'(x)&=\frac{1}{2}x^{-1/2}g_{\lambda,0}(\ln\kappa+\ln x)+x^{-1/2}g_{\lambda,0}'(\ln\kappa+\ln x)
\\
&=\frac{\sqrt{\kappa}}{2}e^{-\frac{1}{2}(\ln\kappa+\ln x)}g_{\lambda,0}(\ln\kappa+\ln x)+\sqrt{\kappa}e^{-\frac{1}{2}(\ln\kappa+\ln x)}g_{\lambda,0}'(\ln\kappa+\ln x)
\\
&\longrightarrow\frac{\sqrt{\kappa}}{2}+\frac{\sqrt{\kappa}}{2}=\sqrt{\kappa}
\end{split}
\end{equation}
as $x\to0$ since $g_{\lambda,0}$ is the regular solution to an equation of the form $g''=[L^2+W]g$ with $L=0+1/2=1/2$.
\QED
\begin{cor}\label{PhiInDomainCor}
For each $\kappa>0$ and $\ell\in\mathbb{N}_0$ the function $\widetilde{f}_{\kappa,\ell}$ satisfies $\xi \widetilde{f}_{\kappa,\ell}\in D(H_{\kappa,\ell})$ where $\xi$ is as described in Section \ref{sec2}.
\end{cor}
\proof
Observe that the function $\xi \widetilde{f}_{\kappa,\ell}-\xi f_{\kappa,\ell}$ is in $C^2(\mathbb{R}_+)$ and has compact support. Hence, it is a standard check that it is also in $D(H_{\kappa,\ell,\text{min}})\subseteq D(H_{\kappa,\ell})$ which by Lemma \ref{PhiInDomain} implies the assertion.
\qed
For later use we need some continuity properties of the regular solutions from Proposition \ref{ExistenceOfSolutions} as a function of the potential $W$ in order to control $g_{\lambda,\ell}$ and $\widetilde{g}_{\lambda,\ell}$ near minus infinity. These will be the last abstract results on regular solutions in the sense of Definition \ref{RegularSolution} presented here, and we do so now to keep the treatment hereof somewhat concise. Note that the use of tildes in Lemma \ref{RegSolutionEstimate} below is \underline{not} related to the use of tildes in other parts of the presentation.
\begin{lemma}\label{RegSolutionEstimate}
Let $W$ and $\widetilde{W}$ be two real-valued continuous potentials satisfying (\ref{IntegrablePotentials}) for all $x\in\mathbb{R}$ and denote by $\widetilde{Q}(x)$ the number corresponding to the one defined in (\ref{IntegrablePotentials}) with $W$ replaced by $\widetilde{W}$. If, for some fixed $L>0$, $g$ and $\widetilde{g}$ are the regular solutions to $g''=[L^2+W]g$ and $\widetilde{g}''=[L^2+\widetilde{W}]\widetilde{g}$ respectively then
\[
\abs{g(x)-\widetilde{g}(x)}\leq e^{Lx}L^{-1}D(x)e^{(Q(x)+\widetilde{Q}(x))/L}
\]
for all $x\in\mathbb{R}$ where
\[
D(x):=\int_{-\infty}^{x}\abs{W(y)-\widetilde{W}(y)}\,dy.
\]
\end{lemma}
\proof
Denoting by $h_i$ the functions from the proof of Proposition \ref{ExistenceOfSolutions} and by $\widetilde{h}_i$ the similar quantities for $\widetilde{W}$ readily see that the bounds
\[
\abs{h_i(x)-\widetilde{h}_i(x)}\leq e^{Lx}L^{-i}D(x)\frac{(Q(x)+\widetilde{Q}(x))^{i-1}}{(i-1)!}
\]
for $i=1,2,\dots$ imply the bound in the lemma. The proof of these is by induction, starting with
\[
\abs{h_1(x)-\widetilde{h}_1(x)}\leq\frac{1}{L}\int_{-\infty}^{x}\sinh\bigl(L(x-y)\bigr)\abs{W(y)-\widetilde{W}(y)}e^{Ly}\,dy\leq e^{Lx}L^{-1}D(x).
\]
Then, for $i=1,2,\dots$,
\begin{align*}
\abs{h_{i+1}&(x)-\widetilde{h}_{i+1}(x)}
\\
&\leq\frac{1}{L}\int_{-\infty}^{x}\sinh\bigl(L(x-y)\bigr)\abs{W(y)h_i(y)-\widetilde{W}(y)\widetilde{h}_i(y)}\,dy
\\
&\leq\frac{1}{L}\int_{-\infty}^{x}\sinh\bigl(L(x-y)\bigr)\bigl[\abs{W(y)-\widetilde{W}(y)}\cdot\abs{h_i(y)}+\abs{\widetilde{W}(y)}\cdot\abs{h_i(y)-\widetilde{h}_i(y)}\bigr]\,dy
\\
&\leq\frac{e^{Lx}}{L}\int_{-\infty}^{x}\abs{W(y)-\widetilde{W}(y)}\frac{Q(y)^i}{L^ii!}+\abs{\widetilde{W}(y)}\cdot D(y)\frac{(Q(y)+\widetilde{Q}(y))^{i-1}}{L^i(i-1)!}\,dy
\\
&\leq\frac{e^{Lx}}{L^{i+1}}D(x)\Bigl[\frac{Q(x)^i}{i!}+\int_{-\infty}^{x}\abs{\widetilde{W}(y)}\frac{(Q(y)+\widetilde{Q}(y))^{i-1}}{(i-1)!}\,dy\Bigr],
\end{align*}
where we used (\ref{ExponentialEstimate}) and the induction hypothesis in the third inequality. Furthermore,
\begin{align*}
\int_{-\infty}^{x}\abs{\widetilde{W}(y)}&\frac{(Q(y)+\widetilde{Q}(y))^{i-1}}{(i-1)!}\,dy
\\
&=\frac{1}{(i-1)!}\sum_{j=0}^{i-1}\binom{i-1}{j}\int_{-\infty}^{x}\abs{\widetilde{W}(y)}Q(y)^j\widetilde{Q}(y)^{i-1-j}\,dy
\\
&\leq\frac{1}{(i-1)!}\sum_{j=0}^{i-1}\binom{i-1}{j}Q(x)^j\int_{-\infty}^{x}\abs{\widetilde{W}(y)}\widetilde{Q}(y)^{i-1-j}\,dy
\\
&=\frac{1}{(i-1)!}\sum_{j=0}^{i-1}\binom{i-1}{j}Q(x)^j\frac{\widetilde{Q}(x)^{i-j}}{i-j}=\frac{1}{i!}\sum_{j=0}^{i-1}\binom{i}{j}Q(x)^j\widetilde{Q}(y)^{i-j}
\\
&=\frac{(Q(x)+\widetilde{Q}(x))^i}{i!}-\frac{Q(x)^i}{i!},
\end{align*}
finishing the induction.
\QED
\subsection{Application of the Liouville-Green approximation}
We now aim for determining parts of the asymptotic behaviour of $g_{\lambda,\ell}$ and $\widetilde{g}_{\lambda,\ell}$ for fixed $\ell$ in the $\lambda\to\infty$ limit away from $x=\pm\infty$, i.e. on the part of the real axis where we cannot take advantage of our knowledge of the asymptotics of $\Phi$. For this we use the Liouville-Green (LG) approximation and in particular the very precise pointwise estimates for this given in Theorem 4 in \cite{Olver}. Slightly reformulating this result to adapt it to our set-up, it reads as presented in Proposition \ref{OlverEst} below. We remind the reader that our conventions for "$\propto$" and "$\lesssim$" are as stated in Subsection \ref{Strategy}.
\begin{pro}[\cite{Olver} Theorem~4]\label{OlverEst}
For $\lambda>0$, let $W_\lambda$ be a $C^2$ and strictly positive potential defined on some finite interval $(x_1(\lambda),x_2(\lambda))$ that might depend on the value of $\lambda$. Now, if $g_\lambda$ is real-valued and solves the equation $g_\lambda''=-\lambda^2W_\lambda g_\lambda$ then
\[
g_\lambda(x)\propto W_\lambda(x)^{-1/4}\Bigl[\cos\Bigl(\lambda\int_{x_1(\lambda)}^{x}W_\lambda^{1/2}\,dy+\theta(\lambda)\Bigr)+\varepsilon_\lambda(x)\Bigr]
\]
for some number $\theta(\lambda)$ with the error estimate
\[
\frac{\abs{\varepsilon_\lambda(x)}}{2}\leq\exp\Bigl(\lambda^{-1}\int_{x_1(\lambda)}^{x_2(\lambda)}W_\lambda^{-1/4}\Big\vert\frac{d^2}{dx^2}\bigl(W_\lambda^{-1/4}\bigr)\Big\vert\,dx\Bigr)-1
\]
for all $x\in(x_1(\lambda),x_2(\lambda))$.
\end{pro}
In order to apply Proposition \ref{OlverEst} we should first of all find an appropriate interval $(x_1(\lambda),x_2(\lambda))$ on which $V_{\lambda,\ell}$ and $\widetilde{V}_{\lambda,\ell}$ are positive -- and then we can hope to be able to control the error estimate in Proposition \ref{OlverEst} on this interval. As a first step towards this notice that due to 2) and 3) in Assumptions \ref{Assumptions} we have
\begin{equation}\label{SimpleBounds1}
b_0e^{(2+\alpha)x}-\lambda^{-2}\bigl(\ell+\frac{1}{2}\bigr)^2\leq V_{\lambda,\ell}(x),\widetilde{V}_{\lambda,\ell}(x)\leq B_0e^{(2+\alpha)x}-\lambda^{-2}\bigl(\ell+\frac{1}{2}\bigr)^2
\end{equation}
for $x\in(-\infty,0)$ and
\begin{equation}\label{SimpleBounds2}
b_\infty e^{(2+\beta)x}-\lambda^{-2}\bigl(\ell+\frac{1}{2}\bigr)^2\leq V_{\lambda,\ell}(x),\widetilde{V}_{\lambda,\ell}(x)\leq B_\infty e^{(2+\beta)x}-\lambda^{-2}\bigl(\ell+\frac{1}{2}\bigr)^2
\end{equation}
for $x\in(0,\infty)$ for some positive constants $b_0,B_0,b_\infty$ and $B_\infty$. In particular this implies that $V_{\lambda,\ell}$ and $\widetilde{V}_{\lambda,\ell}$ are positive on the interval
\[
\bigl(\frac{2\ln(\ell+1/2)-\ln b_0-2\ln\lambda}{2+\alpha},\frac{2\ln(\ell+1/2)-\ln b_\infty-2\ln\lambda}{2+\beta}\bigr),
\]
and initially this could be our guess for where to apply the LG approximation for large $\lambda$. However, we need to be a tiny bit more restrictive than this and take as the interval for our LG approximations
\[
(x_1(\lambda),x_2(\lambda))=\bigl(-\frac{2\ln\lambda}{2+\alpha}+\eta(\lambda),-\frac{2\ln\lambda}{2+\beta}-\eta(\lambda)\bigr)
\]
on which $V_{\lambda,\ell}$ and $\widetilde{V}_{\lambda,\ell}$ are clearly positive for sufficiently large $\lambda$ if $\eta(\lambda)$ tends towards $\infty$ as $\lambda\to\infty$. Of course the interval is empty if $\eta(\lambda)$ tends too fast to infinity. \textbf{We will generally consider only $\eta$ so that this is not the case, in particular so that the endpoint tends towards $-\infty$ and $+\infty$ respectively.} As a first result we have:
\begin{lemma}\label{LGApproximation1}
For any $\ell\in\mathbb{N}_0$ there exists a family $\{\theta_0(\lambda,\ell)\}_{\lambda>0}$ of constants so that
\begin{equation}\label{LGEquation1}
g_{\lambda,\ell}(x)\propto V_{\lambda,\ell}(x)^{-1/4}\Bigl[\cos\Bigl(\lambda\int_{-\infty}^{x}[V_{\lambda,\ell}]_+^{1/2}\,dy+\theta_0(\lambda,\ell)\Bigr)+o_{\lambda\to\infty}(1)\Bigr]
\end{equation}
for $x\in (-\frac{2\ln\lambda}{2+\alpha}+\eta(\lambda),-\frac{2\ln\lambda}{2+\beta}-\eta(\lambda))$ whenever $\eta(\lambda)\to\infty$ as $\lambda\to\infty$. Here, $o_{\lambda\to\infty}(1)$ is uniform in $x$ on this interval.
\end{lemma}
\proof
Note firstly that it suffices to prove the statement in the lemma for any fixed $\eta$ tending towards infinity at infinity which we thus consider in the following. We observe that
\[
V_{\lambda,\ell}(x)<0\quad\text{for}\quad x<\frac{2\ln(\ell+1/2)-\ln B_0-2\ln\lambda}{2+\alpha}
\]
by (\ref{SimpleBounds1}) so that the integral in (\ref{LGEquation1}) is well-defined. Since, moreover, this integral differs from
\[
\int_{-\frac{2\ln\lambda}{2+\alpha}+\eta(\lambda)}^{x}V_{\lambda,\ell}(y)^{1/2}\,dy
\]
only by a constant (in $x$), the result (\ref{LGEquation1}) follows from Proposition \ref{OlverEst} if we manage to show that
\begin{equation}\label{LGErrorEst}
\begin{split}
\lambda^{-1}\int_{-\frac{2\ln\lambda}{2+\alpha}+\eta(\lambda)}^{-\frac{2\ln\lambda}{2+\beta}-\eta(\lambda)}&V_{\lambda,\ell}^{-1/4}\Big\vert\frac{d^2}{dx^2}\bigl(V_{\lambda,\ell}^{-1/4}\bigr)\Big\vert\,dx
\\
&\lesssim\lambda^{-1}\int_{\frac{-2\ln\lambda}{2+\alpha}+\eta(\lambda)}^{\frac{-2\ln\lambda}{2+\beta}-\eta(\lambda)}\frac{\abs{V_{\lambda,\ell}'(x)}^2}{V_{\lambda,\ell}(x)^{5/2}}+\frac{\abs{V_{\lambda,\ell}''(x)}}{V_{\lambda,\ell}(x)^{3/2}}\,dx
\end{split}
\end{equation}
tends towards $0$ as $\lambda\to\infty$. By 5) in Assumptions \ref{Assumptions} we obtain straightforwardly
\[
\abs{V_{\lambda,\ell}'(x)},\abs{V_{\lambda,\ell}''(x)}\lesssim
\begin{dcases*}
e^{(2+\alpha)x} & on $(-\infty,0)$
\\
e^{(2+\beta)x} & on $(0,\infty)$
\end{dcases*}
\]
which we can use to verify the described convergence. In particular, we see that
\begin{align*}
\lambda^{-1}\int_{\frac{-2\ln\lambda}{2+\alpha}+\eta(\lambda)}^{0}&\frac{\abs{V_{\lambda,\ell}'(x)}^2}{V_{\lambda,\ell}(x)^{5/2}}\,dx\lesssim\lambda^{-1}\int_{\frac{-2\ln\lambda}{2+\alpha}+\eta(\lambda)}^{0}\frac{e^{(4+2\alpha)x}}{\bigl[b_0e^{(2+\alpha)x}-\lambda^{-2}\bigl(\ell+\frac{1}{2}\bigr)^2\bigr]^{5/2}}\,dx
\\
&=\lambda^{-1}\int_{\frac{-2\ln\lambda}{2+\alpha}+\eta(\lambda)}^{0}\bigl[b_0e^{\frac{1}{5}(2+\alpha)x}-\lambda^{-2}\bigl(\ell+\frac{1}{2}\bigr)^2e^{-\frac{4}{5}(2+\alpha)x}\bigr]^{-5/2}\,dx
\\
&=\Biggl[\lambda^{-1}\frac{4\lambda^{-2}\bigl(\ell+\frac{1}{2}\bigr)^2-6b_0e^{(2+\alpha)x}}{3(2+\alpha)b_0^2\bigl[b_0e^{(2+\alpha)x}-\lambda^{-2}\bigl(\ell+\frac{1}{2}\bigr)^2\bigr]^{3/2}}\Biggr]^{0}_{\frac{-2\ln\lambda}{2+\alpha}+\eta(\lambda)}
\\
&=\Biggl[\frac{4\bigl(\ell+\frac{1}{2}\bigr)^2-6b_0\lambda^2e^{(2+\alpha)x}}{3(2+\alpha)b_0^2\bigl[b_0\lambda^2e^{(2+\alpha)x}-\bigl(\ell+\frac{1}{2}\bigr)^2\bigr]^{3/2}}\Biggr]^{0}_{\frac{-2\ln\lambda}{2+\alpha}+\eta(\lambda)}\longrightarrow 0
\end{align*}
as $\lambda\to\infty$ by insertion of the limits, as well as
\begin{align*}
\lambda^{-1}\int_{\frac{-2\ln\lambda}{2+\alpha}+\eta(\lambda)}^{0}&\frac{\abs{V_{\lambda,\ell}''(x)}}{V_{\lambda,\ell}(x)^{3/2}}\,dx\lesssim\lambda^{-1}\int_{\frac{-2\ln\lambda}{2+\alpha}+\eta(\lambda)}^{0}\frac{e^{(2+\alpha)x}}{\bigl[b_0e^{(2+\alpha)x}-\lambda^{-2}\bigl(\ell+\frac{1}{2}\bigr)^2\bigr]^{3/2}}\,dx
\\
&=\lambda^{-1}\int_{\frac{-2\ln\lambda}{2+\alpha}+\eta(\lambda)}^{0}\bigl[b_0e^{\frac{1}{3}(2+\alpha)x}-\lambda^{-2}\bigl(\ell+\frac{1}{2}\bigr)^2e^{-\frac{2}{3}(2+\alpha)x}\bigr]^{-3/2}\,dx
\\
&=\Bigl[-2\lambda^{-1}b_0^{-1}(2+\alpha)^{-1}\bigl[b_0e^{(2+\alpha)x}-\lambda^{-2}\bigl(\ell+\frac{1}{2}\bigr)^2\bigr]^{-1/2}\Bigr]^{0}_{\frac{-2\ln\lambda}{2+\alpha}+\eta(\lambda)}
\\
&=\Bigl[-2b_0^{-1}(2+\alpha)^{-1}\bigl[b_0\lambda^2e^{(2+\alpha)x}-\bigl(\ell+\frac{1}{2}\bigr)^2\bigr]^{-1/2}\Bigr]^{0}_{\frac{-2\ln\lambda}{2+\alpha}+\eta(\lambda)}\longrightarrow 0.
\end{align*}
Completely analogously we have
\[
\lambda^{-1}\int_{0}^{\frac{-2\ln\lambda}{2+\beta}-\eta(\lambda)}\frac{\abs{V_{\lambda,\ell}'(x)}^2}{V_{\lambda,\ell}(x)^{5/2}}\,dx\lesssim\Biggl[\frac{4\bigl(\ell+\frac{1}{2}\bigr)^2-6b_\infty\lambda^2e^{(2+\beta)x}}{3(2+\beta)b_\infty^2\bigl[b_\infty\lambda^2e^{(2+\beta)x}-\bigl(\ell+\frac{1}{2}\bigr)^2\bigr]^{3/2}}\Biggr]_{0}^{\frac{-2\ln\lambda}{2+\beta}-\eta(\lambda)}
\]
and
\begin{align*}
\lambda^{-1}\int_{0}^{\frac{-2\ln\lambda}{2+\beta}-\eta(\lambda)}&\frac{\abs{V_{\lambda,\ell}''(x)}}{V_{\lambda,\ell}(x)^{3/2}}\,dx
\\
&\lesssim\Bigl[-2b_\infty^{-1}(2+\beta)^{-1}\bigl[b_\infty\lambda^2e^{(2+\beta)x}-\bigl(\ell+\frac{1}{2}\bigr)^2\bigr]^{-1/2}\Bigr]_{0}^{\frac{-2\ln\lambda}{2+\beta}-\eta(\lambda)},
\end{align*}
both of which tend towards $0$ as $\lambda\to\infty$. This finishes the proof.
\QED
\begin{cor}\label{LGApproximation1Cor}
For any $\ell\in\mathbb{N}_0$ there exists a family $\{\widetilde{\theta}_0(\lambda,\ell)\}_{\lambda>0}$ of constants so that
\begin{equation}\label{LGEquation1Cor}
\widetilde{g}_{\lambda,\ell}(x)\propto \widetilde{V}_{\lambda,\ell}(x)^{-1/4}\Bigl[\cos\Bigl(\lambda\int_{-\infty}^{x}[\widetilde{V}_{\lambda,\ell}]_+^{1/2}\,dy+\widetilde{\theta}_0(\lambda,\ell)\Bigr)+o_{\lambda\to\infty}(1)\Bigr]
\end{equation}
for $x\in (-\frac{2\ln\lambda}{2+\alpha}+\eta(\lambda),-\frac{2\ln\lambda}{2+\beta}-\eta(\lambda))$ whenever $\eta(\lambda)\to\infty$ as $\lambda\to\infty$. Here, $o_{\lambda\to\infty}(1)$ is uniform in $x$ on this interval.
\end{cor}
\proof
Analogously to in the proof of Lemma \ref{LGApproximation1} it suffices to show that the expression (\ref{LGErrorEst}) with $V_{\lambda,\ell}$ replaced by $\widetilde{V}_{\lambda,\ell}$ tends towards zero as $\lambda\to\infty$. Following this proof, it suffices in turn to argue that
\[
\abs{\widetilde{V}_{\lambda,\ell}'(x)},\abs{\widetilde{V}_{\lambda,\ell}''(x)}\lesssim
\begin{dcases*}
e^{(2+\alpha)x} & on $(-\infty,0)$
\\
e^{(2+\beta)x} & on $(0,\infty)$.
\end{dcases*}
\]
Here the bound on $(-\infty,0)$ is as before. The bound on $(0,\infty)$ will follow from proving the equivalent of the second part of 5) in Assumptions \ref{Assumptions} for the modified potentials $\Psi_\lambda$ independently of $\lambda>0$. Using (\ref{ZetaBounds}) we get this from the simple computation
\begin{align*}
\abs{\Psi_\lambda'}=\abs{\zeta_\lambda'(\Phi-\Phi_\infty)+\zeta_\lambda\Phi'+(1-\zeta_\lambda)\Phi_\infty'}&\leq\zeta_\lambda'(\abs{\Phi}+\abs{\Phi_\infty})+\zeta_\lambda\abs{\Phi'}+(1-\zeta_\lambda)\abs{\Phi_\infty'}
\\
&\lesssim x^{\beta-1},
\end{align*}
which hold on $(1,\infty)$, and the completely similar computation yielding $\abs{\Psi_\lambda''}=\cdots\lesssim x^{\beta-2}$ here as needed.
\qed
As a next step we wish to replace the potentials $V_{\lambda,\ell}$ and $\widetilde{V}_{\lambda,\ell}$ with some $\lambda$-independent potentials in the expressions (\ref{LGEquation1}) and (\ref{LGEquation1Cor}) respectively. The factors $V_{\lambda,\ell}(x)^{-1/4}$ and $\widetilde{V}_{\lambda,\ell}(x)^{-1/4}$ are rather easy to rewrite, and hence we focus for the moment our energy on replacing the integrands with some $\lambda$-independent expressions. When doing so we naturally end up changing both the constant terms inside the cosine and the error terms, but this is not important. Knowledge about the result in Lemma \ref{CorrectionToSV} below in particular cases goes back to the early days of quantum mechanics (at least in the non-Langer-transformed set-up). In the very specific case of meromorphic potentials with $\alpha=0$ or $\alpha=-1$ it is essentially Proposition 12 in \cite{ExactWKB} up to a change of variable.
\begin{lemma}\label{CorrectionToSV}
The $\lambda\to\infty$ asymptotics
\[
\lambda\int_{-\infty}^{x}e^{y}\Phi(e^y)^{1/2}\,dy-\lambda\int_{-\infty}^{x}[V_{\lambda,\ell}(y)]^{1/2}_+\,dy=\frac{(2\ell+1)\uppi}{4+2\alpha}+o_{\lambda\to\infty}(1)
\]
holds uniformly for $x\in(\frac{-2\ln\lambda}{2+\alpha}+\eta(\lambda),\frac{-2\ln\lambda}{2+\beta}-\eta(\lambda))$ if $\eta(\lambda)$ tends to infinity as $\lambda\to\infty$.
\end{lemma}
\proof
Notice that it clearly suffices to show the result in the case where $\eta(\lambda)\to\infty$ arbitrarily slowly for $\lambda\to\infty$ which we thus assume. We observe that
\begin{align*}
\abs{\lambda e^y\Phi(e^y)^{1/2}-\lambda V_{\lambda,\ell}(y)^{1/2}}&\leq\lambda^{-1}\bigl(\ell+\frac{1}{2}\bigr)^2e^{-y}\Phi(e^y)^{-1/2}
\\
&\lesssim
\begin{dcases*}
\lambda^{-1}\bigl(\ell+\frac{1}{2}\bigr)^2e^{-\frac{2+\alpha}{2}y} &on $(-\frac{2\ln\lambda}{2+\alpha}+\eta(\lambda),0)$
\\
\lambda^{-1}\bigl(\ell+\frac{1}{2}\bigr)^2e^{-\frac{2+\beta}{2}y} &on $(0,-\frac{2\ln\lambda}{2+\beta}-\eta(\lambda))$
\end{dcases*}
\end{align*}
and that a simple insertion of this in the integral yields
\[
\int_{-\frac{2\ln\lambda}{2+\alpha}+\eta(\lambda)}^{-\frac{2\ln\lambda}{2+\beta}-\eta(\lambda)}\abs{\lambda e^{y}\Phi(e^y)^{1/2}-\lambda V_{\lambda,\ell}(y)^{1/2}}\,dy\longrightarrow0
\]
as $\lambda\to\infty$ as long as $\eta(\lambda)\to\infty$ as $\lambda\to\infty$. Hence, we only need to prove
\begin{equation}\label{CorrectionToSVProof}
\begin{split}
\lambda\int_{-\infty}^{-\frac{2\ln\lambda}{2+\alpha}+\eta(\lambda)}e^{y}\Phi(e^y)^{1/2}\,dy-\lambda\int_{-\infty}^{-\frac{2\ln\lambda}{2+\alpha}+\eta(\lambda)}&[V_{\lambda,\ell}(y)]^{1/2}_+\,dy
\\
&=\frac{(2\ell+1)\uppi}{4+2\alpha}+o_{\lambda\to\infty}(1)
\end{split}
\end{equation}
to have shown the full statement of the lemma. The idea is from this point to approximate $\Phi(e^y)$ by $C_0e^{\alpha y}$ everywhere. For the first term in (\ref{CorrectionToSVProof}) this gives the error
\begin{align*}
\lambda\int_{-\infty}^{-\frac{2\ln\lambda}{2+\alpha}+\eta(\lambda)}e^{y}\abs{\Phi(e^y)&^{1/2}-C_0^{1/2}e^{\frac{\alpha}{2}y}}\,dy
\\
&=\lambda\int_{-\infty}^{-\frac{2\ln\lambda}{2+\alpha}+\eta(\lambda)}e^{(1+\frac{\alpha}{2})y}\abs{e^{-\frac{\alpha}{2}y}\Phi(e^y)^{1/2}-C_0^{1/2}}\,dy
\\
&\lesssim e^{(1+\frac{\alpha}{2})\eta(\lambda)}\sup_{y\leq-\frac{2\ln\lambda}{2+\alpha}+\eta(\lambda)}\abs{e^{-\frac{\alpha}{2}y}\Phi(e^y)^{1/2}-C_0^{1/2}}\longrightarrow 0
\end{align*}
if $\eta$ tends towards $\infty$ sufficiently slowly. Next step is to replace $V_{\lambda,\ell}(y)$ by $C_0e^{(2+\alpha)x}-\lambda^{-2}(\ell+1/2)^2$ in the second term of (\ref{CorrectionToSVProof}). For this we use the general inequality
\[
\big\vert [u]_+^{1/2}-[v]_+^{1/2}\big\vert\leq\abs{u-v}^{1/2}
\]
for real numbers $u$ and $v$ together with
\[
\bigl\vert V_{\lambda,\ell}(y)+\lambda^{-2}\bigl(\ell+\frac{1}{2}\bigr)^2-C_0e^{(2+\alpha)y}\bigr\vert=e^{(2+\alpha)y}\abs{e^{-\alpha y}\Phi(e^y)-C_0}
\]
to conclude that
\begin{align*}
\int_{-\infty}^{-\frac{2\ln\lambda}{2+\alpha}+\eta(\lambda)}\bigl\vert\lambda[V_{\lambda,\ell}(y)]^{1/2}_+-&\lambda\bigl[C_0e^{(2+\alpha)y}-\lambda^{-2}\bigl(\ell+\frac{1}{2}\bigr)^2\bigr]_+^{1/2}\bigr\vert\,dy
\\
&\leq\lambda\int_{-\infty}^{-\frac{2\ln\lambda}{2+\alpha}+\eta(\lambda)}e^{(1+\frac{\alpha}{2})y}\abs{e^{-\alpha y}\Phi(e^y)-C_0}^{1/2}\,dy
\\
&\lesssim e^{(1+\frac{\alpha}{2})\eta(\lambda)}\sup_{y\leq-\frac{2\ln\lambda}{2+\alpha}+\eta(\lambda)}\abs{e^{-\alpha y}\Phi(e^y)-C_0}^{1/2}\longrightarrow 0
\end{align*}
as before. This means that we obtain
\begin{align*}
\lambda\int_{-\infty}^{-\frac{2\ln\lambda}{2+\alpha}+\eta(\lambda)}&e^{y}\Phi(e^y)^{1/2}\,dy-\lambda\int_{-\infty}^{-\frac{2\ln\lambda}{2+\alpha}+\eta(\lambda)}[V_{\lambda,\ell}(y)]^{1/2}_+\,dy
\\
&\hspace{-1.9cm}=\lambda\int_{-\infty}^{-\frac{2\ln\lambda}{2+\alpha}+\eta(\lambda)}C_0^{1/2}e^{(1+\frac{\alpha}{2})y}\,dy-\lambda\int_{-\infty}^{-\frac{2\ln\lambda}{2+\alpha}+\eta(\lambda)}\bigl[C_0e^{(2+\alpha)y}-\lambda^{-2}\bigl(\ell+\frac{1}{2}\bigr)^2\bigr]_+^{1/2}\,dy
\\
&\hspace{8cm}+o_{\lambda\to\infty}(1),
\end{align*}
and an explicit calculation of the right hand side here is all there is left to do in order to prove the lemma. Here, the first term clearly equals
\[
C_0^{1/2}\Bigl(1+\frac{\alpha}{2}\Bigr)^{-1}e^{(1+\frac{\alpha}{2})\eta(\lambda)}=:T_\lambda
\]
which tends towards $+\infty$ as $\lambda\to\infty$, and the second term equals
\begin{align*}
\Bigl(T_\lambda^2-\bigl(1+\frac{\alpha}{2}\bigr)^{-2}\bigl(\ell+\frac{1}{2}\bigr)^2\Bigr)^{1/2}&-\frac{2\ell+1}{2+\alpha}\tan^{-1}\Bigl(\bigl(\ell+\frac{1}{2}\bigr)^{-1}\Bigl(\bigl(1+\frac{\alpha}{2}\Bigr)^2T_\lambda^2-\bigl(\ell+\frac{1}{2}\bigr)^2\Bigr)^{1/2}\Bigr)
\\
&=\bigl(T_\lambda+o_{\lambda\to\infty}(1)\bigr)-\frac{2\ell+1}{2+\alpha}\bigl(\frac{\uppi}{2}+o_{\lambda\to\infty}(1)\bigr).
\end{align*}
Finally, subtracting these yields the claimed result.
\QED
\begin{cor}\label{CorrectionToSVCor}
The $\lambda\to\infty$ asymptotics
\[
\lambda\int_{-\infty}^{x}e^{y}\Phi(e^y)^{1/2}\,dy-\lambda\int_{-\infty}^{x}[\widetilde{V}_{\lambda,\ell}(y)]^{1/2}_+\,dy=\frac{(2\ell+1)\uppi}{4+2\alpha}+o_{\lambda\to\infty}(1)
\]
holds uniformly for $x\in(\frac{-2\ln\lambda}{2+\alpha}+\eta(\lambda),\frac{-2\ln\lambda}{2+\beta}-\eta(\lambda))$ if $\eta(\lambda)$ tends to infinity sufficiently slowly as $\lambda\to\infty$.
\end{cor}
\proof
Observe that, for $x\in(\frac{-2\ln\lambda}{2+\alpha}+\eta(\lambda),\frac{-2\ln\lambda}{2+\beta}-\eta(\lambda))$,
\begin{align*}
\Bigl\vert\lambda\int_{-\infty}^{x}&[V_{\lambda,\ell}(y)]^{1/2}_+\,dy-\lambda\int_{-\infty}^{x}[\widetilde{V}_{\lambda,\ell}(y)]^{1/2}_+\,dy\Bigr\vert\leq\lambda\int_{-\infty}^{x} \!\!\! \big\vert[V_{\lambda,\ell}(y)]^{1/2}_+-[\widetilde{V}_{\lambda,\ell}(y)]^{1/2}_+\big\vert\,dy
\\
&\leq\lambda\int_{-\infty}^{x}\vert V_{\lambda,\ell}(y)-\widetilde{V}_{\lambda,\ell}(y)\vert^{1/2}\,dy=\lambda\int_{-\infty}^{x}e^{y}\vert \Phi(e^y)-\Psi_\lambda(e^y)\vert^{1/2}\,dy
\\
&\leq\lambda\int_{-\frac{2\ln\lambda}{2+\beta}-3\eta(\lambda)}^{-\frac{2\ln\lambda}{2+\beta}-\eta(\lambda)}e^{\frac{2+\beta}{2}y}\vert e^{-\beta y}\Phi(e^y)-C_\infty\vert^{1/2}\,dy
\\
&\lesssim (e^{-\frac{6+3\beta}{2}\eta(\lambda)}-e^{-\frac{2+\beta}{2}\eta(\lambda)})\sup_{y\geq-\frac{2\ln\lambda}{2+\beta}-3\eta(\lambda)}\vert e^{-\beta y}\Phi(e^y)-C_\infty\vert^{1/2}\longrightarrow0
\end{align*}
as $\lambda\to\infty$ if $\eta(\lambda)\to\infty$ sufficiently slowly. Thus, the assertion follows from Lemma \ref{CorrectionToSV}.
\qed
Collecting the pieces in the present subsection, we can now obtain the desired relation between the LG approximation and the solutions $g_{\lambda,\ell}$ and $\widetilde{g}_{\lambda,\ell}$.
\begin{pro}\label{LGApproximation2}
For any $\ell\in\mathbb{N}_0$ there exists a family $\{\theta(\lambda,\ell)\}_{\lambda>0}$ of constants so that
\begin{equation}\label{LGEquation2}
e^{x/2}\Phi(e^x)^{1/4}g_{\lambda,\ell}(x)\propto\cos\Bigl(\lambda\int_{-\infty}^{x}e^y\Phi(e^y)^{1/2}\,dy+\theta(\lambda,\ell)\Bigr)+o_{\lambda\to\infty}(1)
\end{equation}
for $x\in (-\frac{2\ln\lambda}{2+\alpha}+\eta(\lambda),-\frac{2\ln\lambda}{2+\beta}-\eta(\lambda))$ whenever $\eta$ tends to infinity at infinity. Here, $o_{\lambda\to\infty}(1)$ is uniform in $x$ on this interval. In particular, if $\eta$ tends to infinity sufficiently slowly,
\begin{equation}\label{LGEquation3}
e^{\frac{2+\alpha}{4}x}g_{\lambda,\ell}(x)\propto\cos\Bigl(\frac{2C_0^{1/2}\lambda}{2+\alpha}e^{(1+\frac{\alpha}{2})x}+\theta(\lambda,\ell)\Bigr)+o_{\lambda\to\infty}(1)
\end{equation}
for $x\in (-\frac{2\ln\lambda}{2+\alpha}+\eta(\lambda),-\frac{2\ln\lambda}{2+\alpha}+2\eta(\lambda))$ and
\begin{equation}\label{LGEquation4}
e^{\frac{2+\beta}{4}x}g_{\lambda,\ell}(x)\propto\cos\Bigl(\int_{0}^{\infty}\Phi_\kappa(y)^{1/2}\,dy+\frac{2C_\infty^{1/2}\lambda}{2+\beta}e^{(1+\frac{\beta}{2})x}+\theta(\lambda,\ell)\Bigr)+o_{\lambda\to\infty}(1)
\end{equation}
for $x\in (-\frac{2\ln\lambda}{2+\beta}-2\eta(\lambda),-\frac{2\ln\lambda}{2+\beta}-\eta(\lambda))$.
\end{pro}
\proof
The asymptotics in (\ref{LGEquation2}) is basically a consequence of Lemmas \ref{LGApproximation1} and \ref{CorrectionToSV} as soon as one realizes that
\[
V_{\lambda,\ell}(x)^{-1/4}=e^{-x/2}\Phi(e^x)^{-1/4}\bigl(1+o_{\lambda\to\infty}(1)\bigr)
\]
uniformly on the interval. This can be seen for example by examining the last term in
\[
\frac{V_{\lambda,\ell}(x)}{e^{2x}\Phi(e^x)}=1-\lambda^{-2}e^{-2x}\Phi(e^x)^{-1}\bigl(\ell+\frac{1}{2}\bigr)^2
\]
for $x\in(-\frac{2\ln\lambda}{2+\alpha}+\eta(\lambda),-\frac{2\ln\lambda}{2+\beta}-\eta(\lambda))$.

We observe further that
\[
e^{-x/2}\Phi(e^x)^{-1/4}\propto e^{-\frac{2+\alpha}{4}x}(1+o_{\lambda\to\infty}(1))\;\;\text{and}\;\; e^{-x/2}\Phi(e^x)^{-1/4}\propto e^{-\frac{2+\beta}{4}x}(1+o_{\lambda\to\infty}(1))
\]
uniformly on $(-\frac{2\ln\lambda}{2+\alpha}+\eta(\lambda),-\frac{2\ln\lambda}{2+\alpha}+2\eta(\lambda))$ and $(-\frac{2\ln\lambda}{2+\beta}-2\eta(\lambda),-\frac{2\ln\lambda}{2+\beta}-\eta(\lambda))$ respectively. For proving (\ref{LGEquation3}) and (\ref{LGEquation4}) it thus remains only to argue that, assuming $\eta$ tends to infinity sufficiently slowly,
\begin{equation}\label{AsympOfIntegral1}
\lambda\int_{-\infty}^{x}e^y\Phi(e^y)^{1/2}\,dy=\frac{2C_0^{1/2}\lambda}{2+\alpha}e^{(1+\frac{\alpha}{2})x}+o_{\lambda\to\infty}(1)
\end{equation}
and
\begin{equation}\label{AsympOfIntegral2}
\lambda\int_{-\infty}^{x}e^y\Phi(e^y)^{1/2}\,dy=\int_{0}^{\infty}\Phi_\kappa(y)^{1/2}\,dy+\frac{2C_\infty^{1/2}\lambda}{2+\beta}e^{(1+\frac{\beta}{2})x}+o_{\lambda\to\infty}(1)
\end{equation}
uniformly for $x<2\eta(\lambda)-2\ln\lambda/(2+\alpha)$ and $x>-2\eta(\lambda)-2\ln\lambda/(2+\beta)$ respectively. Since (\ref{AsympOfIntegral1}) was more or less proved during the proof of Lemma \ref{CorrectionToSV}, we focus our attention on (\ref{AsympOfIntegral2}). On the interval of interest we have by the change of variables $y\to e^y$ the following estimates:
\begin{align*}
\Bigl\vert\int_{0}^{\infty}&\Phi_\kappa(y)^{1/2}\,dy-\lambda\int_{-\infty}^{x}e^y\Phi(e^y)^{1/2}\,dy+\frac{2C_\infty^{1/2}\lambda}{2+\beta}e^{(1+\frac{\beta}{2})x}\Bigr\vert
\\
&=\Bigl\vert\lambda\int_{x}^{\infty}e^y\Phi(e^y)^{1/2}\,dy-\lambda\int_{x}^{\infty}C_\infty^{1/2}e^{(1+\frac{\beta}{2})y}\,dy\Bigr\vert
\\
&\leq\lambda\int_{x}^{\infty}e^{(1+\frac{\beta}{2})y}\abs{e^{-\frac{\beta}{2}y}\Phi(e^y)^{1/2}-C_\infty^{1/2}}\,dy\lesssim\lambda e^{(1+\frac{\beta}{2})x}\sup_{y\geq x}\abs{e^{-\frac{\beta}{2}y}\Phi(e^y)^{1/2}-C_\infty^{1/2}}
\\
&\leq e^{-(2+\beta)\eta(\lambda)}\sup_{y\geq -\frac{2\ln\lambda}{2+\beta}-2\eta(\lambda)}\abs{e^{-\frac{\beta}{2}y}\Phi(e^y)^{1/2}-C_\infty^{1/2}}\longrightarrow0,
\end{align*}
where the convergence holds as long as $\eta(\lambda)$ tends towards $\infty$ sufficiently slowly as $\lambda\to\infty$. This finishes the proof.
\QED
\begin{cor}\label{LGApproximation2Cor}
For any $\ell\in\mathbb{N}_0$ there exists a family $\{\widetilde{\theta}(\lambda,\ell)\}_{\lambda>0}$ of constants so that
\begin{equation}\label{LGEquation2Cor}
e^{x/2}\Phi(e^x)^{1/4}\widetilde{g}_{\lambda,\ell}(x)\propto\cos\Bigl(\lambda\int_{-\infty}^{x}e^y\Phi(e^y)^{1/2}\,dy+\widetilde{\theta}(\lambda,\ell)\Bigr)+o_{\lambda\to\infty}(1)
\end{equation}
for $x\in (-\frac{2\ln\lambda}{2+\alpha}+\eta(\lambda),-\frac{2\ln\lambda}{2+\beta}-\eta(\lambda))$ whenever $\eta$ tends to infinity sufficiently slowly at infinity. Here, $o_{\lambda\to\infty}(1)$ is uniform in $x$ on this interval. In particular,
\begin{equation}\label{LGEquation3Cor}
e^{\frac{2+\alpha}{4}x}\widetilde{g}_{\lambda,\ell}(x)\propto\cos\Bigl(\frac{2C_0^{1/2}\lambda}{2+\alpha}e^{(1+\frac{\alpha}{2})x}+\widetilde{\theta}(\lambda,\ell)\Bigr)+o_{\lambda\to\infty}(1)
\end{equation}
for $x\in (-\frac{2\ln\lambda}{2+\alpha}+\eta(\lambda),-\frac{2\ln\lambda}{2+\alpha}+2\eta(\lambda))$ and
\begin{equation}\label{LGEquation4Cor}
e^{\frac{2+\beta}{4}x}\widetilde{g}_{\lambda,\ell}(x)\propto\cos\Bigl(\int_{0}^{\infty}\Phi_\kappa(y)^{1/2}\,dy+\frac{2C_\infty^{1/2}\lambda}{2+\beta}e^{(1+\frac{\beta}{2})x}+\widetilde{\theta}(\lambda,\ell)\Bigr)+o_{\lambda\to\infty}(1)
\end{equation}
for $x\in (-\frac{2\ln\lambda}{2+\beta}-2\eta(\lambda),-\frac{2\ln\lambda}{2+\beta}-\eta(\lambda))$.
\end{cor}
\proof
Similar to above, the asymptotics in (\ref{LGEquation2Cor}) follows from Corollaries \ref{LGApproximation1Cor} and \ref{CorrectionToSVCor} as soon as one realizes that
\begin{equation}\label{VTildeApproxEq}
\widetilde{V}_{\lambda,\ell}(x)^{-1/4}=e^{-x/2}\Phi(e^x)^{-1/4}\bigl(1+o_{\lambda\to\infty}(1)\bigr)
\end{equation}
uniformly on $(-\frac{2\ln\lambda}{2+\alpha}+\eta(\lambda),-\frac{2\ln\lambda}{2+\beta}-\eta(\lambda))$. For this we consider
\[
\frac{\widetilde{V}_{\lambda,\ell}(x)}{e^{2x}\Phi(e^x)}=\frac{\Psi_\lambda(e^x)}{\Phi(e^x)}-\lambda^{-2}e^{-2x}\Phi(e^x)^{-1}\bigl(\ell+\frac{1}{2}\bigr)^2=\frac{\Psi_\lambda(e^x)}{\Phi(e^x)}+o_{\lambda\to\infty}(1)
\]
and notice that $\Psi_\lambda(e^x)/\Phi(e^x)=1$ for $x<-\frac{2\ln\lambda}{2+\beta}-3\eta(\lambda)$ as well as
\[
\Big\vert\frac{\Psi_\lambda(e^x)}{\Phi(e^x)}-1\Big\vert\leq\Big\vert\frac{\Phi_\infty(e^x)}{\Phi(e^x)}-1\Big\vert\longrightarrow0
\]
uniformly on $[-\frac{2\ln\lambda}{2+\beta}-3\eta(\lambda),-\frac{2\ln\lambda}{2+\beta}-\eta(\lambda))$. This yields (\ref{VTildeApproxEq}) and thus, in turn, (\ref{LGEquation2Cor}). From here one can follow the proof of Proposition \ref{LGApproximation2} exactly to arrive at (\ref{LGEquation3Cor}) and (\ref{LGEquation4Cor}).
\QED
\subsection{Gluing together the solutions}
From Proposition \ref{LGApproximation2} and Corollary \ref{LGApproximation2Cor} we need to control the constants $\theta(\lambda,\ell)$ and $\widetilde{\theta}(\lambda,\ell)$ and to examine the behaviour of $\widetilde{g}_{\lambda,\ell}$ further to the right where the LG-approximation no longer works. For both these tasks we use the asymptotic behaviour of $\Phi$, firstly combined with Lemma \ref{RegSolutionEstimate}.
\begin{lemma}\label{ThetaAsymptotics}
Let $\{\theta(\lambda,\ell)\}_{\lambda>0}$ be families of constants so that (\ref{LGEquation2}), (\ref{LGEquation3}) and (\ref{LGEquation4}) hold true for intervals as in Proposition \ref{LGApproximation2} (for some $\eta$). Let likewise $\{\widetilde{\theta}(\lambda,\ell)\}_{\lambda>0}$ be families of constants so that (\ref{LGEquation2Cor}), (\ref{LGEquation3Cor}) and (\ref{LGEquation4Cor}) hold true for intervals as in Corollary~\ref{LGApproximation2Cor} (for some $\eta$). Then these constants satisfy
\[
\theta(\lambda,\ell),\widetilde{\theta}(\lambda,\ell)\longrightarrow-\frac{(2\ell+1)\uppi}{4+2\alpha}-\frac{\uppi}{4}\quad\textup{(mod $\;\uppi$)}
\]
as $\lambda\to\infty$.
\end{lemma}
\proof
We observe initially that we can freely assume that $\eta$ tends arbitrarily slowly towards infinity at infinity as this does not affect the values of the $\theta(\lambda,\ell)$'s and the $\widetilde{\theta}(\lambda,\ell)$'s.

The main part of the proof is to show the convergence for the $\theta(\lambda,\ell)$'s. For this, the idea is to consider the regular solutions $h_{\lambda,\ell}$ to the equations $h_{\lambda,\ell}''=[L^2+W_{\lambda}]h_{\lambda,\ell}$ with $L=\ell+1/2$ and $W_{\lambda}(x)=-\lambda^2 C_0e^{(2+\alpha)x}$ and compare this to $g_{\lambda,\ell}$ on $(-\frac{2\ln\lambda}{2+\alpha}+\eta(\lambda),-\frac{2\ln\lambda}{2+\alpha}+2\eta(\lambda))$. By the help of Lemma \ref{RegSolutionEstimate} it will be possible to infer the needed asymptotics of $\theta(\lambda,\ell)$ from this comparison.

In the spirit of Lemma \ref{RegSolutionEstimate} we define thus the quantities
\[
\quad Q_\lambda(x)=\lambda^2\int_{-\infty}^{x}e^{2y}\Phi(e^y)\,dy\quad\text{and}\quad\widetilde{Q}_\lambda(x)=C_0\lambda^2\int_{-\infty}^{x}e^{(2+\alpha)y}\,dy,
\]
noticing that $Q_\lambda(x)\lesssim\widetilde{Q}_\lambda(x)\lesssim\lambda^2 e^{(2+\alpha)x}$ for $x<0$, and
\[
D_\lambda(x)=\lambda^2\int_{-\infty}^{x}\abs{e^{2y}\Phi(e^{y})-C_0e^{(2+\alpha)y}}\,dy.
\]
Now the space of solutions to the equation that $h_{\lambda,\ell}$ solves is spanned by the real-valued functions
\[
J_{\frac{2\ell+1}{2+\alpha}}\Bigl(\frac{2C_0^{1/2}\lambda}{2+\alpha}e^{\frac{2+\alpha}{2}x}\Bigr)\qquad\text{and}\qquad Y_{\frac{2\ell+1}{2+\alpha}}\Bigl(\frac{2C_0^{1/2}\lambda}{2+\alpha}e^{\frac{2+\alpha}{2}x}\Bigr),
\]
and from the characterization of the regular solution from Proposition \ref{ExistenceOfSolutions} together with the asymptotic behaviour of the Bessel function at the origin (cf. \cite{AS} p.360) one can conclude that
\begin{equation}\label{ExpForH}
h_{\lambda,\ell}(x)=A(\ell,\alpha,C_0)\lambda^{-\frac{2\ell+1}{2+\alpha}}J_{\frac{2\ell+1}{2+\alpha}}\Bigl(\frac{2C_0^{1/2}\lambda}{2+\alpha}e^{\frac{2+\alpha}{2}x}\Bigr).
\end{equation}
with $A(\ell,\alpha,C_0)$ a constant independent of $\lambda$. Hence, when looking at $x\in (-\frac{2\ln\lambda}{2+\alpha}+\eta(\lambda),-\frac{2\ln\lambda}{2+\alpha}+2\eta(\lambda))$, we can use the asymptotics of the Bessel function at infinity (cf. \cite{AS} p.364) to see that
\begin{equation}\label{AsymptoticOscillation}
\lambda^{\frac{2\ell+1}{2+\alpha}+\frac{1}{2}}e^{\frac{2+\alpha}{4}x}h_{\lambda,\ell}(x)=A'(\ell,\alpha,C_0)\cos\Bigl(\frac{2C_0^{1/2}\lambda}{2+\alpha}e^{\frac{2+\alpha}{2}x}-\frac{(2\ell+1)\uppi}{4+2\alpha}-\frac{\uppi}{4}\Bigr)+o_{\lambda\to\infty}(1)
\end{equation}
uniformly on this interval with $A'(\ell,\alpha,C_0)$ a constant independent of $\lambda$. Applying Lemma \ref{RegSolutionEstimate} we obtain also, for all $x<-\frac{2\ln\lambda}{2+\alpha}+2\eta(\lambda)$, the key inequalities
\begin{align*}
\lambda^{\frac{2\ell+1}{2+\alpha}+\frac{1}{2}}e^{\frac{2+\alpha}{4}x}&\abs{g_{\lambda,\ell}(x)-h_{\lambda,\ell}(x)}\lesssim\lambda^{\frac{2\ell+1}{2+\alpha}+\frac{1}{2}}e^{\frac{2+\alpha}{4}x}e^{(\ell+\frac{1}{2})x}D_\lambda(x)e^{c'(Q_\lambda(x)+\widetilde{Q}_\lambda(x))}
\\
&\leq\exp\Bigl(\frac{4\ell+4+\alpha}{4+2\alpha}\ln\lambda+\frac{4\ell+4+\alpha}{4}x\Bigr)D_\lambda(x)e^{c''\widetilde{Q}_\lambda(x)}
\\
&\leq e^{(2\ell+2+\frac{\alpha}{2})\eta(\lambda)}D_\lambda\bigl(\frac{-2\ln\lambda}{2+\alpha}+2\eta(\lambda)\bigr)e^{c''\widetilde{Q}_\lambda\bigl(\frac{-2\ln\lambda}{2+\alpha}+2\eta(\lambda)\bigr)}
\\
&\leq e^{(2\ell+2+\frac{\alpha}{2})\eta(\lambda)}e^{c'''e^{2(2+\alpha)\eta(\lambda)}}D_\lambda\bigl(\frac{-2\ln\lambda}{2+\alpha}+2\eta(\lambda)\bigr)
\end{align*}
for some $\ell$-depending constants $c'$, $c''$ and $c'''$ where
\begin{align*}
D_\lambda\bigl(\frac{-2\ln\lambda}{2+\alpha}+2\eta(\lambda)\bigr)&=\lambda^2\int_{-\infty}^{\frac{-2\ln\lambda}{2+\alpha}+2\eta(\lambda)}e^{(2+\alpha)y}\abs{e^{-\alpha y}\Phi(e^y)-C_0}\,dy
\\
&\lesssim e^{2(2+\alpha)\eta(\lambda)}\sup_{y\leq-\frac{2\ln\lambda}{2+\alpha}+2\eta(\lambda)}\abs{e^{-\alpha y}\Phi(e^y)-C_0}.
\end{align*}
If $\eta(\lambda)\to\infty$ sufficiently slowly for $\lambda\to\infty$ then this shows that
\begin{equation}\label{GApproxH}
\lambda^{\frac{2\ell+1}{2+\alpha}+\frac{1}{2}}e^{\frac{2+\alpha}{4}x}\abs{g_{\lambda,\ell}(x)-h_{\lambda,\ell}(x)}\longrightarrow0
\end{equation}
uniformly on this interval. We conclude from this and the expression (\ref{AsymptoticOscillation}) for $h_{\lambda,\ell}$ that (\ref{AsymptoticOscillation}) remains true when replacing $h_{\lambda,\ell}$ by $g_{\lambda,\ell}$. Comparing with (\ref{LGEquation3}),
\begin{align*}
\cos\Bigl(\frac{2C_0^{1/2}\lambda}{2+\alpha}e^{(1+\frac{\alpha}{2})x}+&\theta(\lambda,\ell)\Bigr)+o_{\lambda\to\infty}(1)
\\
&\propto\;\;\cos\Bigl(\frac{2C_0^{1/2}\lambda}{2+\alpha}e^{(1+\frac{\alpha}{2})x}-\frac{(2\ell+1)\uppi}{4+2\alpha}-\frac{\uppi}{4}\Bigr)+o_{\lambda\to\infty}(1)
\end{align*}
for $x\in (-\frac{2\ln\lambda}{2+\alpha}+\eta(\lambda),-\frac{2\ln\lambda}{2+\alpha}+2\eta(\lambda))$ with uniform errors. Since $\lambda e^{(1+\frac{\alpha}{2})x}$ ranges over arbitrarily large values for $x$ varying in this interval, it must be the case that the constant terms inside the cosines agree asymptotically modulo $\uppi$, proving exactly the desired convergence of the $\theta(\lambda,\ell)$'s.

Notice finally that we have throughout the proof considered $g_{\lambda,\ell}(x)$ only for $x<-\frac{2\ln\lambda}{2+\alpha}+2\eta(\lambda)$, i.e. on a set where it agrees with $\widetilde{g}_{\lambda,\ell}(x)$. Hence, (\ref{LGEquation3Cor}) tells us that also
\begin{align*}
\cos\Bigl(\frac{2C_0^{1/2}\lambda}{2+\alpha}e^{(1+\frac{\alpha}{2})x}+&\widetilde{\theta}(\lambda,\ell)\Bigr)+o_{\lambda\to\infty}(1)
\\
&\propto\;\;\cos\Bigl(\frac{2C_0^{1/2}\lambda}{2+\alpha}e^{(1+\frac{\alpha}{2})x}-\frac{(2\ell+1)\uppi}{4+2\alpha}-\frac{\uppi}{4}\Bigr)+o_{\lambda\to\infty}(1)
\end{align*}
for $x\in (-\frac{2\ln\lambda}{2+\alpha}+\eta(\lambda),-\frac{2\ln\lambda}{2+\alpha}+2\eta(\lambda))$ with uniform errors -- from which the desired convergence of the $\widetilde{\theta}(\lambda,\ell)$'s follows.
\qed
Combining Proposition \ref{LGApproximation2} and Corollary \ref{LGApproximation2Cor} (in particular (\ref{LGEquation2}), (\ref{LGEquation2Cor}) and (\ref{LGEquation4Cor})) with Lemma \ref{ThetaAsymptotics} we obtain directly:
\begin{cor}\label{LGApproximation3}
For any $\ell\in\mathbb{N}_0$,
\begin{equation}\label{LGEquationFinal}
e^{x/2}\Phi(e^x)^{1/4}g_{\lambda,\ell}(x)\propto\cos\Bigl(\lambda\int_{-\infty}^{x}e^y\Phi(e^y)^{1/2}\,dy-\frac{(2\ell+1)\uppi}{4+2\alpha}-\frac{\uppi}{4}\Bigr)+o_{\lambda\to\infty}(1)
\end{equation}
and
\begin{equation}\label{LGEquationFinalCor}
e^{x/2}\Phi(e^x)^{1/4}\widetilde{g}_{\lambda,\ell}(x)\propto\cos\Bigl(\lambda\int_{-\infty}^{x}e^y\Phi(e^y)^{1/2}\,dy-\frac{(2\ell+1)\uppi}{4+2\alpha}-\frac{\uppi}{4}\Bigr)+o_{\lambda\to\infty}(1)
\end{equation}
for $x\in (-\frac{2\ln\lambda}{2+\alpha}+\eta(\lambda),-\frac{2\ln\lambda}{2+\beta}-\eta(\lambda))$ if $\eta$ tends to infinity sufficiently slowly at infinity and where $o_{\lambda\to\infty}(1)$ is uniform in $x$ on this interval. Further,
\[
e^{\frac{2+\beta}{4}x}\widetilde{g}_{\lambda,\ell}(x)\propto\cos\Bigl(\int_{0}^{\infty}\Phi_\kappa(y)^{1/2}\,dy+\frac{2C_\infty^{1/2}\lambda}{2+\beta}e^{(1+\frac{\beta}{2})x}-\frac{(2\ell+1)\uppi}{4+2\alpha}-\frac{\uppi}{4}\Bigr)+o_{\lambda\to\infty}(1)
\]
for $x\in (-\frac{2\ln\lambda}{2+\beta}-2\eta(\lambda),-\frac{2\ln\lambda}{2+\beta}-\eta(\lambda))$ where $o_{\lambda\to\infty}(1)$ is uniform in $x$ on this interval.
\end{cor}
We can now use (\ref{LGEquationFinal}) to obtain a general result on the asymptotic behaviour of solutions to (\ref{StandardWKBEq}) for large $\lambda$. Notice that taking $w_{\lambda,\ell}(x):=\sqrt{x}g_{\lambda,\ell}(\ln x)=\kappa^{1/2}f_{\lambda,\ell}(\kappa^{-1}x)$, the equation in (\ref{StandardWKBEq}) is satisfied. These functions satisfy additionally
\[
x^{-\ell-1}w_{\lambda,\ell}(x)=x^{-(\ell+\frac{1}{2})}g_{\lambda,\ell}(\ln x)\longrightarrow 1
\]
as $x\to0$ since $g_{\lambda,\ell}$ is the regular solution to an equation with $L=\ell+1/2$. Moreover, it can be seen similarly to in (\ref{fIsRegular}) that $x^{-\ell}w'_{\lambda,\ell}(x)\to\ell+1$ as $x\to0$. As explained in Section \ref{sec2} the $w_{\lambda,\ell}$'s must be the unique solutions to (\ref{StandardWKBEq}) satisfying these boundary conditions -- for $\ell=0$ by \cite{JanD2}, Proposition 2.5, and for $\ell=1,2,\dots$ since we are then in the limit point case meaning that there is additionally a solution not in $L^2$ near the origin. By insertion in (\ref{LGEquationFinal}) we learn that
\[
w_{\lambda,\ell}(x)\propto \Phi(x)^{-1/4}\Bigl(\cos\Bigl(\lambda\int_{-\infty}^{\ln x}e^y\Phi(e^y)^{1/2}\,dy-\frac{(2\ell+1)\uppi}{4+2\alpha}-\frac{\uppi}{4}\Bigr)+o_{\lambda\to\infty}(1)\Bigr)
\]
for $x\in(\lambda^{\frac{-2}{2+\alpha}}e^{\eta(\lambda)},\lambda^{\frac{-2}{2+\beta}}e^{-\eta(\lambda)})$ with $o_{\lambda\to\infty}(1)$ uniform on this interval. Applying Lemma \ref{CorrectionToSV} and changing variable in the integral now yields straightforwardly:
\begin{thm}\label{JWKBLanger}
Assume that $\Phi$ is a potential satisfying the Assumptions \ref{Assumptions} and let $\ell\in\mathbb{N}_0$ be given. Then, for each $\lambda>0$, there exists a unique solution $w_{\lambda,\ell}$ to the equation (\ref{StandardWKBEq}) with $x^{-\ell-1}w_{\lambda,\ell}(x)\to1$ and $x^{-\ell}w'_{\lambda,\ell}(x)\to\ell+1$ as $x\to0$. Whenever $x_-(\lambda)$ and $x_+(\lambda)$ satisfy $\lambda^{\frac{2}{2+\alpha}}x_-(\lambda)\to\infty$ and $\lambda^{\frac{2}{2+\beta}}x_+(\lambda)\to0$ as $\lambda\to\infty$, these solutions have the form
\[
w_{\lambda,\ell}(x)\propto \Phi(x)^{-1/4}\Bigl(\cos\Bigl(\lambda\int_{0}^{x}\Bigl[\Phi(y)-\frac{\bigl(\ell+\frac{1}{2}\bigr)^2}{\lambda^2y^2}\Bigr]_+^{1/2}\,dy-\frac{\uppi}{4}\Bigr)+o_{\lambda\to\infty}(1)\Bigr)
\]
for $x\in(x_-(\lambda),x_+(\lambda))$ with $o_{\lambda\to\infty}(1)$ uniform on this interval.
\end{thm}
\subsection{Proof of Theorem \ref{MainResult}}\label{ProofSubSec}
We are now in a position to prove our main result, choosing from this point onwards $\eta$ to satisfy the properties listed in Subsection \ref{Strategy}. The proof is split into two parts: Part 1 which formalizes more or less the discussion in Subsection \ref{Strategy}, and Part 2 which takes care of a technical issue of bounding certain normalized solutions near the \vspace{.2cm} origin.
\\
\textbf{\underline{Part 1:} }Assume that $\kappa_n\to\infty$ and (\ref{MainResultEquation}) as $n\to\infty$ and consider also the numbers $\lambda_n=\kappa_n^{-(2+\beta)/2}\to\infty$. From the discussion in Subsection~\ref{subsec3.1} it is sufficient to construct $\phi_{n,\ell}$ satisfying the conditions in Proposition \ref{GraphLemma2}.

We take as our candidates for these functions $\phi_{n,\ell}:=c_n\xi\widetilde{f}_{\kappa_n,\ell}$ where $\widetilde{f}_{\kappa,\ell}(x)=\sqrt{x}\,\widetilde{g}_{\lambda,\ell}(\ln\kappa+\ln x)$ as in (\ref{fDef}) and $c_n\neq0$ are constants to be determined below. Now $\phi_{n,\ell}\in D(H_{\kappa_n,\ell})$ by Corollary \ref{PhiInDomainCor}. Also, according to Lemma \ref{PhiInDomain},
\begin{equation}\label{PhiSolvesEquation}
H_{\kappa_n,\ell}\phi_{n,\ell}(x)=0\qquad\text{for}\quad x<e^{-3\eta(\lambda_n)}
\end{equation}
since $\phi_{n,\ell}=c_nf_{\kappa_n,\ell}$ on this interval. We now proceed to verify the two convergence properties listed in Proposition \ref{GraphLemma2} for our choice of $\phi_{n,\ell}$'s.

For the first convergence we take a closer look at the $\widetilde{g}_{\lambda_n,\ell}$'s on the intervals $(-\frac{2\ln\lambda_n}{2+\beta}-2\eta(\lambda_n),\infty)$. As they solve the equations $\widetilde{g}_{\lambda_n,\ell}''(x)=[(\ell+1/2)^2-\lambda_n^2C_\infty e^{(2+\beta)x}]\widetilde{g}_{\lambda_n,\ell}(x)$ here and are real-valued, they must be on the form
\[
\widetilde{g}_{\lambda_n,\ell}(x)=\frac{1}{c_n}\Bigl[\cos\theta_{n,\ell}\cdot J_{\frac{2\ell+1}{2+\beta}}\Bigl(\frac{-2C_\infty^{1/2}\lambda_n}{2+\beta}e^{\frac{2+\beta}{2}x}\Bigr)+\sin\theta_{n,\ell}\cdot Y_{\frac{2\ell+1}{2+\beta}}\Bigl(\frac{-2C_\infty^{1/2}\lambda_n}{2+\beta}e^{\frac{2+\beta}{2}x}\Bigr)\Bigr]
\]
for some numbers $\theta_{n,\ell}$ which are fixed modulo $\uppi$ and some\footnote{Note that the $c_n$'s may also depend on $\ell$. Since this will not be important in our arguments, we suppress it in the notation.} real number $c_n\neq0$. These are the $c_n$'s we use to define our $\phi_{n,\ell}$'s. In particular this means that
\begin{equation}\label{PhiAsymptotics}
\phi_{n,\ell}=\xi(\cos\theta_{n,\ell} F_{\beta,C_\infty,\ell}+\sin\theta_{n,\ell} G_{\beta,C_\infty,\ell})
\end{equation}
on $(e^{-2\eta(\lambda_n)},\infty)$. Thus, it will prove the first convergence property in Proposition~\ref{GraphLemma2} if we manage to show that
\begin{equation}\label{ThetaEllConvergence}
\theta_{n,\ell}\longrightarrow\tau\uppi-\frac{(2\ell+1)\uppi}{4+2\alpha}-\frac{(2\ell+1)\uppi}{4+2\beta}-\frac{\uppi}{2}=:\theta_\ell\quad\text{(mod $\;\uppi$)}\quad\text{\emph{as}}\quad n\to\infty
\end{equation}
and
\begin{equation}\label{UniformBoundNear0}
\text{\emph{for each $\ell\in\mathbb{N}_0$, $\phi_{n,\ell}$ tends uniformly towards $0$ on $(0,e^{-2\eta(\lambda_n)})$}}.
\end{equation}
For the former we notice that on $(e^{-2\eta(\lambda_n)},e^{-\eta(\lambda_n)})$ we have on the one hand, by (\ref{PhiAsymptotics}) and the asymptotics of the Bessel-functions (cf. \cite{AS} p.364),
\[
x^{\beta/4}\phi_{n,\ell}(x)\propto\cos\Bigl(\theta_{n,\ell}+\frac{2C_\infty^{1/2}}{2+\beta}x^{\frac{2+\beta}{2}}+\frac{(2\ell+1)\uppi}{4+2\beta}+\frac{\uppi}{4}\Bigr)+o_{n\to\infty}(1)
\]
uniformly on the interval. On the other hand, by using the definition of $\phi_{n,\ell}$, Corollary \ref{LGApproximation3} and (\ref{MainResultEquation}) we have, also uniformly on this interval,
\begin{align*}
x^{\beta/4}\phi_{n,\ell}(x)&\propto\cos\Bigl(\int_{0}^{\infty}\Phi_{\kappa_n}^{1/2}\,dy+\frac{2C_\infty^{1/2}}{2+\beta}x^{\frac{2+\beta}{2}}-\frac{(2\ell+1)\uppi}{4+2\alpha}-\frac{\uppi}{4}\Bigr)+o_{n\to\infty}(1)
\\
&=\cos\Bigl(\tau\uppi+\frac{2C_\infty^{1/2}}{2+\beta}x^{\frac{2+\beta}{2}}-\frac{(2\ell+1)\uppi}{4+2\alpha}-\frac{\uppi}{4}\Bigr)+o_{n\to\infty}(1),
\end{align*}
and, as $x^{\frac{2+\beta}{2}}$ ranges over arbitrarily large values for $x\in(e^{-2\eta(\lambda_n)},e^{-\eta(\lambda_n)})$, we conclude that
\[
\theta_{n,\ell}+\frac{(2\ell+1)\uppi}{4+2\beta}+\frac{\uppi}{4}\longrightarrow\tau\uppi-\frac{(2\ell+1)\uppi}{4+2\alpha}-\frac{\uppi}{4}
\]
modulo $\uppi$ from which (\ref{ThetaEllConvergence}) follows. The property (\ref{UniformBoundNear0}) is proved below in Part 2 of the proof.

In order to verify the second convergence property in Proposition~\ref{GraphLemma2} notice that this amounts to proving
\[
\int_{0}^{\infty}\abs{H_{\kappa_n,\ell}\phi_{n,\ell}-H_{\infty,\ell,\theta_\ell}(\xi(\cos\theta_\ell F_{\beta,C_\infty,\ell}+\sin\theta_\ell G_{\beta,C_\infty,\ell}))}^2\,dx\longrightarrow0
\]
as $n\to\infty$. On $(0,e^{-3\eta(\lambda_n)})$ and $(2,\infty)$ the integrand is $0$ by (\ref{PhiSolvesEquation}) and the definition of $\xi$ so we need not worry about these parts of $\mathbb{R}_+$. Further, on $(1,2)$ we get by (\ref{ThetaEllConvergence}) uniform convergence of $(\cos\theta_{n,\ell}F_{\beta,C_\infty,\ell}+\sin\theta_{n,\ell}G_{\beta,C_\infty,\ell})^{(p)}$ towards $(\cos\theta_\ell F_{\beta,C_\infty,\ell}+\sin\theta_\ell G_{\beta,C_\infty,\ell})^{(p)}$ for $p=0,1,2$, and -- together with the uniform convergence of $\Phi_{\kappa_n}$ towards $\Phi_\infty$ on this interval -- this rather straightforwardly takes care of the $L^2$-convergence here. We are thus left with the task of estimating
\begin{align*}
\int_{e^{-3\eta(\lambda_n)}}^{1}\abs{&H_{\kappa_n,\ell}\phi_{n,\ell}-H_{\infty,\ell,\theta_\ell}(\xi(\cos\theta_\ell F_{\beta,C_\infty,\ell}+\sin\theta_\ell G_{\beta,C_\infty,\ell}))}^2\,dx
\\
&=\int_{e^{-3\eta(\lambda_n)}}^{1}\abs{H_{\kappa_n,\ell}\phi_{n,\ell}}^2\,dx\leq\int_{e^{-3\eta(\lambda_n)}}^{1}\abs{\phi_{n,\ell}}^2\cdot\abs{\Phi_{\kappa_n}-\Phi_\infty}^2\,dx
\\
&\lesssim\int_{e^{-3\eta(\lambda_n)}}^{\infty}\abs{\Phi_{\kappa_n}-\Phi_\infty}^2\,dx=\kappa_n^{-2\beta}\int_{e^{-3\eta(\lambda_n)}}^{\infty}\abs{\Phi(\kappa_nx)-C_\infty(\kappa_nx)^\beta}^2\,dx
\\
&=\kappa_n^{-2\beta-1}\int_{\kappa_ne^{-3\eta(\lambda_n)}}^{\infty}\abs{\Phi(x)-C_\infty x^\beta}^2\,dx
\\
&=\kappa_n^{-2\beta-1}\int_{\kappa_ne^{-3\eta(\lambda_n)}}^{\infty}x^{2\beta}\abs{x^{-\beta}\Phi(x)-C_\infty}^2\,dx
\\
&\lesssim e^{-3(2\beta+1)\eta(\lambda_n)}\sup_{x\geq\kappa_ne^{-3\eta(\lambda_n)}}\abs{x^{-\beta}\Phi(x)-C_\infty}^2\longrightarrow0
\end{align*}
as long as $\eta(\lambda_n)\to\infty$ sufficiently slowly as $n\to\infty$, where we have used (\ref{PhiAsymptotics}) and (\ref{UniformBoundNear0}) along the way. This proves the second convergence property in Proposition \ref{GraphLemma2} and thus proves the main theorem -- up to proving the property (\ref{UniformBoundNear0}).
\\ \\
\textbf{\underline{Part 2:} } We now focus on proving (\ref{UniformBoundNear0}). Firstly note that for $x<e^{-2\eta(\lambda_n)}$ we have
\[
\phi_{n,\ell}(x)=c_n\widetilde{f}_{\kappa_n,\ell}(x)=c_n\sqrt{x}\widetilde{g}_{\lambda_n,\ell}(\ln\kappa_n+\ln x)=c_n\kappa_n^{-1/2}e^{\frac{\ln\kappa_n+\ln x}{2}}\widetilde{g}_{\lambda_n,\ell}(\ln\kappa_n+\ln x)
\]
so that proving (\ref{UniformBoundNear0}) actually amounts to arguing that $c_n\kappa_n^{-1/2}e^{x/2}\widetilde{g}_{\lambda_n,\ell}(x)$ converges uniformly towards $0$ on $(-\infty,\ln\kappa_n-2\eta(\lambda_n))=(-\infty,\frac{-2\ln\lambda_n}{2+\beta}-2\eta(\lambda_n))$. We denote this property by (\ref{UniformBoundNear0})'.

Now in order to prove (\ref{UniformBoundNear0})' we observe that on $(-\frac{2\ln\lambda_n}{2+\beta}-2\eta(\lambda_n),-\frac{2\ln\lambda_n}{2+\beta}-\eta(\lambda_n))$, by the definition of $c_n$ and the asymptotics of the Bessel functions,
\begin{align*}
c_n\lambda_n^{1/2}e^{\frac{2+\beta}{4}x}&\widetilde{g}_{\lambda_n,\ell}(x)
\\
&=B(\beta,C_\infty)\cos\Bigl(\theta_{n,\ell}+\frac{2C_\infty^{1/2}\lambda_n}{2+\beta}e^{\frac{2+\beta}{2}x}+\frac{(2\ell+1)\uppi}{4+2\beta}+\frac{\uppi}{4}\Bigr)+o_{n\to\infty}(1)
\end{align*}
where $B(\beta,C_\infty)$ is a constant independent of $n$. Since $e^{\frac{2+\beta}{4}x}=C_\infty^{-1/4}e^{x/2}\Phi(e^x)^{1/4}(1+o_{n\to\infty}(1))$ on this interval (where, as usual, $\lambda_ne^{\frac{2+\beta}{2}x}$ ranges over arbitrarily large values), (\ref{LGEquationFinalCor}) tells us that 
\begin{align*}
c_n\lambda_n^{1/2}e^{x/2}&\Phi(e^x)^{1/4}\widetilde{g}_{\lambda_n,\ell}(x)
\\
&=B'(\beta,C_\infty)\cos\Bigl(\lambda\int_{-\infty}^{x}e^y\Phi(e^y)^{1/2}\,dy-\frac{(2\ell+1)\uppi}{4+2\alpha}-\frac{\uppi}{4}\Bigr)+o_{n\to\infty}(1)
\end{align*}
uniformly on the (larger) interval $(\frac{-2\ln\lambda_n}{2+\alpha}+\eta(\lambda_n),\frac{-2\ln\lambda_n}{2+\beta}-\eta(\lambda_n))$, where $B'(\beta,C_\infty)$ is a constant independent of $n$. Note that a priori the implicit constant factor in (\ref{LGEquationFinalCor}) may depend on $\lambda=\lambda_n$, but we can conclude that the constant $B'(\beta,C_\infty)$ does not. In particular, on this larger interval,
\begin{align*}
\abs{c_n}\kappa_n^{-1/2}e^{x/2}\abs{\widetilde{g}_{\lambda_n,\ell}(x)}\lesssim\kappa_n^{-1/2}\lambda_n^{-1/2}\Phi(e^x)^{-1/4}&\lesssim\kappa_n^{\beta/4}\cdot\max\{e^{-\frac{\alpha}{4}x},e^{-\frac{\beta}{4}x}\}
\\
&\leq\max\{\lambda_n^\gamma,e^{\frac{\beta}{4}\eta(\lambda_n)}\}\longrightarrow0,
\end{align*}
where "$\lesssim$" means less than up to a constant independent of $n$, and
\[
\gamma:=\frac{\alpha}{2\alpha+4}-\frac{\beta}{2\beta+4}<0.
\]
Thus, we need now only prove (\ref{UniformBoundNear0})' on the remaining part of the interval, i.e. on $(-\infty,\frac{-2\ln\lambda_n}{2+\alpha}+\eta(\lambda_n))$.

For this we need some refined knowledge about the constant $c_n$. To obtain this we note that on this interval $\widetilde{g}_{\lambda_n,\ell}=g_{\lambda_n,\ell}$ and focus our attention on the interval $(\frac{-2\ln\lambda_n}{2+\alpha}+\eta(\lambda_n),\frac{-2\ln\lambda_n}{2+\alpha}+2\eta(\lambda_n))$ for the moment. Here, we have basically just argued that
\[
\abs{c_n}\kappa_n^{-1/2}e^{\frac{2+\alpha}{4}x}\abs{g_{\lambda_n,\ell}(x)}\lesssim\abs{c_n}\kappa_n^{-1/2}e^{x/2}\Phi(e^x)^{1/4}\abs{\widetilde{g}_{\lambda_n,\ell}(x)}\lesssim\kappa_n^{-1/2}\lambda_n^{-1/2}=\kappa_n^{\beta/4},
\]
but on the other hand (\ref{AsymptoticOscillation}) and (\ref{GApproxH}) show that the function $\lambda_n^{\frac{2\ell+1}{2+\alpha}+\frac{1}{2}}e^{\frac{2+\alpha}{4}x}\abs{g_{\lambda_n,\ell}(x)}$ does not converge towards $0$ on this interval. We can thus conclude that we have $\abs{c_n}\kappa_n^{-1/2}\lesssim\kappa_n^{\beta/4}\lambda_n^{\frac{2\ell+1}{2+\alpha}+\frac{1}{2}}$, and consequently that
\begin{equation}\label{LastUniformBound}
\begin{split}
\abs{c_n}\kappa_n^{-1/2}&e^{x/2}\abs{g_{\lambda_n,\ell}(x)}
\\
&\lesssim\kappa_n^{\beta/4}\lambda_n^{\frac{2\ell+1}{2+\alpha}+\frac{1}{2}}e^{x/2}\abs{h_{\lambda_n,\ell}(x)}+\kappa_n^{\beta/4}\lambda_n^{\frac{2\ell+1}{2+\alpha}+\frac{1}{2}}e^{x/2}\abs{g_{\lambda_n,\ell}(x)-h_{\lambda_n,\ell}(x)}
\end{split}
\end{equation}
uniformly on all of $(-\infty,\frac{-2\ln\lambda_n}{2+\alpha}+\eta(\lambda_n))$ where $h_{\lambda_n,\ell}$ is as in the proof of Lemma \ref{ThetaAsymptotics}. Here, the first term on the right hand side is (cf. (\ref{ExpForH}))
\begin{equation}\label{BesselPartOfEtimate}
\kappa^{\beta/4}e^{-\frac{\alpha}{4}x}A(\ell,\alpha,C_0)\lambda_n^{1/2}e^{\frac{2+\alpha}{4}x}\Bigl\vert J_{\frac{2\ell+1}{2+\alpha}}\Bigl(\frac{2C_0^{1/2}\lambda_n}{2+\alpha}e^{\frac{2+\alpha}{2}x}\Bigr)\Bigr\vert
\end{equation}
which we claim tends uniformly towards $0$ on $(-\infty,\frac{-2\ln\lambda_n}{2+\alpha}+\eta(\lambda_n))$. Indeed, to realize that this is the case on $(\frac{-2\ln\lambda_n}{2+\alpha},\frac{-2\ln\lambda_n}{2+\alpha}+\eta(\lambda_n))$ we can use the facts that $\kappa_n^{\beta/4}e^{-\frac{\alpha}{4}x}\leq\max\{\lambda_n^\gamma,\kappa_n^{\beta/4}\}\to0$ here and that $y\mapsto\sqrt{y}J_\nu(y)$ is uniformly bounded on $\mathbb{R}_+$ for $\nu>0$. On $(-\infty,\frac{-2\ln\lambda_n}{2+\alpha})$ we see that the expression inside the Bessel function is bounded and hence so is the Bessel function part of (\ref{BesselPartOfEtimate}). Noticing that $\kappa^{\beta/4}\lambda_n^{1/2}=\kappa_n^{-1/2}\to0$ it follows easily that all of (\ref{BesselPartOfEtimate}) tends towards $0$ uniformly here as well. Finally, we claim that also the last term on the right hand side in (\ref{LastUniformBound}) tends uniformly towards $0$ on $(-\infty,\frac{-2\ln\lambda_n}{2+\alpha}+\eta(\lambda_n))$. On $(\frac{-2\ln\lambda_n}{2+\alpha},\frac{-2\ln\lambda_n}{2+\alpha}+\eta(\lambda_n))$ we can as before use the fact that $\kappa^{\beta/4}e^{-\frac{\alpha}{4}x}\to0$ uniformly and (\ref{GApproxH}) to realize that this is the case here. On $(-\infty,\frac{-2\ln\lambda_n}{2+\alpha})$ we need some calculations using the notation from the proof of Lemma \ref{ThetaAsymptotics}: It is an easy check that here $Q_{\lambda_n}$ and $\widetilde{Q}_{\lambda_n}$ (and hence $D_{\lambda_n}$) are uniformly bounded as functions of $x$ and $n$. Thus, we get -- similarly to in the proof of Lemma \ref{ThetaAsymptotics} -- the inequalities
\begin{align*}
    \kappa_n^{\beta/4}\lambda_n^{\frac{2\ell+1}{2+\alpha}+\frac{1}{2}}e^{x/2}\abs{g_{\lambda_n,\ell}(x)-h_{\lambda_n,\ell}(x)}&\lesssim\kappa_n^{\beta/4}\lambda_n^{\frac{2\ell+1}{2+\alpha}+\frac{1}{2}}e^{(\ell+1)x}
\\
&\leq\lambda_n^{-\frac{\beta}{2\beta+4}+\frac{2\ell+1}{2+\alpha}+\frac{1}{2}-\frac{2\ell+2}{2+\alpha}}=\lambda_n^\gamma\longrightarrow0
\end{align*}
here. Combining these uniform convergences with the bound (\ref{LastUniformBound}) we have managed to prove (\ref{UniformBoundNear0})', finishing the proof of Theorem \ref{MainResult}.
\QED
\section{No norm resolvent convergence}\label{sec4}
We argue in this section that one cannot generally strengthen the convergence of the operators in the "if"-part of the result in Theorem \ref{MainResultTF} to be in the norm resolvent sense. Recall that norm resolvent convergence of a sequence $\{A_n\}_{n=1}^\infty$ of self-adjoint operators towards $A$ simply means norm convergence of the resolvent operators $(A_n-i)^{-1}\to(A-i)^{-1}$ or, equivalently, that $h(A_n)\to h(A)$ in norm for any continuous function $h$ on $\mathbb{R}$ that tends towards $0$ at $\pm\infty$.

More precisely, we will prove that the conditions $Z_n\to\infty$ and (\ref{MainResultEquationTF}) are not sufficient to conclude that $H_{Z_n}^\text{TF}$ is convergent in this stronger sense. For this we use the natural notational convention of adding "TF" to any of the operators in Section \ref{sec2} to indicate that it is defined by the Thomas-Fermi potential $\Phi_1^{\text{TF}}$ (or $\Phi_\infty^{\text{TF}}$). Firstly, we need some intermediate results.
\begin{lemma}\label{NoNormLemma}
Consider the set-up from Section \ref{sec2}. Then the following is true:
\begin{itemize}
\item[a)] For each $\mu<0$ and $\ell\in\mathbb{N}_0$ there exists a $\theta(\ell,\mu)\in[0,\uppi)$ so that $\mu$ is an eigenvalue for $H_{\infty,\ell,\theta(\ell,\mu)}^{\textup{TF}}$,
\item[b)] If $H_{Z_n}^{\textup{TF}}\to H_{\infty,\{\theta_\ell\}_{\ell
=1}^\infty}^{\textup{TF}}$ in the strong resolvent sense as $n\to\infty$ for some sequences $\{Z_n\}_{n=1}^\infty\subseteq\mathbb{R}_+$ and $\{\theta_\ell\}_{\ell=0}^\infty\subseteq[0,\uppi)$ then also $H_{Z_n,\ell}^{\textup{TF}}\to H_{\infty,\ell,\theta_\ell}^{\textup{TF}}$ in the strong resolvent sense as $n\to\infty$ for all $\ell\in\mathbb{N}_0$,
\item[c)] There exist sequences $\{Z_\ell\}_{\ell=1}^\infty\subseteq\mathbb{N}$ and $\{\mu_\ell\}_{\ell=1}^\infty\subseteq(-\infty,0)$ so that $\mu_\ell\in\sigma(H_{Z_\ell,\ell}^{\textup{TF}})$ for each $\ell=1,2,3,\dots$ and so that $Z_\ell\to\infty$ and $\mu_\ell\to-1$ as $\ell\to\infty$. 
\end{itemize}
\end{lemma}
\proof
$a$) For this part we claim that it suffices to find a real-valued solution $f\in L^2(\mathbb{R}_+)$ to the equation
\begin{equation}\label{muEquation}
-f''(x)+\bigl[\frac{\ell(\ell+1)}{x^2}-\Phi_\infty^{\textup{TF}}(x)\bigr]f(x)=\mu f(x).
\end{equation}
To see this we recall the fundamental structure of the extensions of $H_{\infty,\ell,\text{min}}^{\text{TF}}$ which relies on the fact that we are (cf. Section \ref{sec2}) in the limit circle case. For the details see Appendix A in \cite{JanD1}. One has
\begin{equation}\label{HMaxDomain}
\begin{split}
\{\phi\in L^2(\mathbb{R}_+)\vert-\phi''+[\ell(\ell+1)x^{-2}-\Phi_\infty^{\text{TF}}]\phi\in L^2(\mathbb{R}_+)\}&=D((H_{\infty,\ell,\text{min}}^{\textup{TF}})^*)
\\
&\hspace{-1cm}=D(H_{\infty,\ell,\text{min}}^{\textup{TF}})\oplus\mathbb{C}\xi f_1\oplus\mathbb{C}\xi f_2
\end{split}
\end{equation}
with $f_1$ and $f_2$ two linearly independent solutions to the equation $f''=[\ell(\ell+1)x^{-2}-\Phi_\infty^{\text{TF}}]f$. We observe that clearly a solution $f\in L^2(\mathbb{R}_+)$ to (\ref{muEquation}) will be in the domain (\ref{HMaxDomain}), and consequently it must be in $D(H_{\infty,\ell,\text{min}}^{\textup{TF}})\oplus\mathbb{C}\xi\widetilde{f}$ for some real-valued\footnote{The fact that $\widetilde{f}$ must be real-valued relies crucially on the facts that $f$ itself is real-valued and that the operator commutes with complex conjugation.} solution $\widetilde{f}$ to $\widetilde{f}''=[\ell(\ell+1)x^{-2}-\Phi_\infty^{\text{TF}}]\widetilde{f}$. But any domain on this form is the domain of one of the self-adjoint extensions $H_{\infty,\ell,\theta_\ell}^{\textup{TF}}$ of $H_{\infty,\ell,\text{min}}^{\textup{TF}}$ from Section \ref{sec2}. Thus, the assertion follows from (\ref{muEquation}).

Now to find a real-valued $f\in L^2(\mathbb{R}_+)$ solving (\ref{muEquation}) we apply Proposition \ref{ExistenceOfSolutions} with $L=\sqrt{-\mu}$ and $W(x)=\ell(\ell+1)x^{-2}-\Phi_\infty^\text{TF}(-x)$ to get a solution $g$ to 
\[
g''(x)=\bigl[-\mu+\frac{\ell(\ell+1)}{x^2}-\Phi_\infty^{\textup{TF}}(-x)\bigr]g(x)
\]
with $e^{-\sqrt{-\mu}x}g(x)\to1$ as $x\to-\infty$. Considering $f(x):=g(-x)$ we get in this way a real-valued solution to (\ref{muEquation}) satisfying $f\in L^2((1,\infty))$. Moreover, it is a general consequence (cf. \cite{ReedSimon2} Theorem X.6) of being in the limit circle case that all solutions $f$ to (\ref{muEquation}) for \emph{\textbf{any}} $\mu\in \mathbb{C}$ are $L^2$ near the origin, say on $(0,1)$. Hence, we have found the desired $f$.

$b$) This is simply an exercise in digesting the definitions of the operators in Section \ref{sec2}. It is easily verified that
\[
(H_{Z_n,\ell}^{\text{TF}}+i)^{-1}\phi\otimes Y_\ell^0=(\widetilde{H}_{Z_n}^{\text{TF}}+i)^{-1}(\phi\otimes Y_\ell^0)
\]
for any $\phi\in L^2(\mathbb{R}_+)$, and similarly for the operators defining infinite atoms. Consequently, if $H_{Z_n}^{\text{TF}}\to H_{\infty,\{\theta_\ell\}_{\ell=1}^\infty}^{\textup{TF}}$ in the strong resolvent sense (and since we have chosen the spherical harmonics to be normalized in $L^2(S^2)$),
\begin{align*}
\Vert\,(H_{Z_n,\ell}^{\text{TF}}&+i)^{-1}\phi-(H_{\infty,\ell,\theta_\ell}^{\text{TF}}+i)^{-1}\phi\,\Vert
\\
&=\Vert\,(\widetilde{H}_{Z_n}^{\text{TF}}+i)^{-1}(\phi\otimes Y_\ell^0)-(\widetilde{H}_{\infty,\{\theta_\ell\}_{\ell=0}^\infty}^{\text{TF}}+i)^{-1}(\phi\otimes Y_\ell^0)\,\Vert\longrightarrow0
\end{align*}
as $n\to\infty$ for any $\phi\in L^2(\mathbb{R}_+)$. This proves the claimed strong resolvent convergence.

$c$) We will now combine the results from ($a$) and ($b$), and will prove by induction that there exist natural numbers $Z_1<Z_2<Z_3<\cdots$ so that
\begin{equation}\label{muConvergence}
 \text{there exists $\mu_\ell\in\sigma(H_{Z_\ell,\ell}^{\text{TF}})$ so that $\vert\,\mu_\ell+1\,\vert<\ell^{-1}$}
\end{equation}
for all $\ell$. To this end we fix $\ell$ and show that we can have (\ref{muConvergence}) for arbitrarily large $Z_\ell$'s.

Choose $\theta_\ell'$ so that
\begin{equation}\label{ThetaPrime}
\theta_\ell'+\frac{\ell\uppi}{2}+\frac{\uppi}{4}=\theta(\ell,-1)
\end{equation}
modulo $\uppi$ where $\theta(\ell,-1)$ is the number from ($a$). Then choose $Z_{\ell,n}$ to be the number from the right-hand side of (\ref{IntegerCharge}) with $\tau=\theta_\ell'/\uppi$ for each $n=1,2,\dots$ so that in particular $Z_{\ell,n}\to\infty$ as $n\to\infty$. From Theorem \ref{MainResultTF} we learn that then $H_{Z_{\ell,n}}^{\text{TF}}\to H_{\infty,\{\theta_m\}_{m=0}^\infty}^{\text{TF}}$ in the strong resolvent sense as $n\to\infty$ with $\{\theta_m\}_{m=0}^\infty$ defined by
\[
\theta_m=\theta_\ell'+\frac{m\uppi}{2}+\frac{\uppi}{4}.
\]
Thus, by ($b$) and (\ref{ThetaPrime}), $H_{Z_{\ell,n},\ell}^{\text{TF}}\to H_{\infty,\ell,\theta(\ell,-1)}^{\text{TF}}$ in the strong resolvent sense as $n\to\infty$. Moreover, since $-1\in\sigma(H_{\infty,\ell,\theta(\ell,-1)}^{\text{TF}})$, we find by the general concept of spectral exclusion under strong resolvent convergence (cf. \cite{ReedSimon1} Theorem VIII.24) that there are numbers $\mu_{\ell,n}\in\sigma(H_{Z_{\ell,n},\ell}^{\text{TF}})\subseteq\mathbb{R}$ so that $\mu_{\ell,n}\to-1$ as $n\to\infty$. Now choosing $n_0$ sufficiently large we can achieve both $Z_{\ell,n_0}>Z_{\ell-1}$ and $\vert\,\mu_{\ell,n_0}+1\,\vert<\ell^{-1}$, and by setting $Z_\ell:=Z_{\ell,n_0}$ and $\mu_\ell:=\mu_{\ell,n_0}$ we complete the proof.
\QED
\begin{lemma}\label{EigenVectors}
Let $\{Z_\ell\}_{\ell=1}^{\infty}$ and $\{\mu_\ell\}_{\ell=1}^{\infty}$ be sequences as in Lemma \ref{NoNormLemma}($c$). For each $\ell=1,2,3,\dots$ the number $\mu_\ell$ is an eigenvalue of $\widetilde{H}_{Z_\ell}^{\text{TF}}$, and there exists a non-zero eigenvector
\[
\phi_\ell\in D(\widetilde{H}_{Z_\ell}^{\textup{TF}})\cap[L^2(\mathbb{R}_+)\otimes\operatorname{span}Y_\ell^0]=:D(\widetilde{H}_{Z_\ell}^{\textup{TF}})\cap\mathcal{V}_\ell
\]
so that $\widetilde{H}_{Z_\ell}^{\textup{TF}}\phi_\ell=\mu_\ell\phi_\ell$.
\end{lemma}
\proof
We recall firstly that since $H_{Z_\ell}^{\textup{TF}}$ is the closure of $-\Delta-\Phi_{Z_\ell}^{\text{TF}}$ on $C_0^\infty(\mathbb{R}^3)$ and $\Phi_{Z_\ell}^{\text{TF}}\in L^2(\mathbb{R}^3)$ it is a standard consequence of Weyl's Theorem that $\sigma_{\operatorname{ess}}(\widetilde{H}_{Z_\ell}^{\text{TF}})=\sigma_{\operatorname{ess}}(H_{Z_\ell}^{\text{TF}})=[0,\infty)$.

Fix $\ell$. As $\mu_\ell\in \sigma(H_{Z_\ell,\ell}^{\text{TF}})$ there exists a sequence $\{\psi_{\ell,n}\}_{n=1}^{\infty}\subseteq D(H_{Z_\ell,\ell}^{\text{TF}})$ so that $\Vert\,\psi_{\ell,n}\,\Vert=1$ and $\Vert\,H_{Z_\ell,\ell}^{\text{TF}}\psi_{\ell,n}-\mu_\ell\psi_{\ell,n}\,\Vert\longrightarrow0$ as $n\to\infty$. Letting
\[
\{\phi_{\ell,n}\}_{n=1}^{\infty}:=\bigl\{\psi_{\ell,n}\otimes Y_\ell^0\bigr\}_{n=1}^{\infty}\subseteq D(H_{Z_\ell,\ell}^{\text{TF}})\otimes\operatorname{span}Y_\ell^0\subseteq D(\widetilde{H}_{Z_\ell}^{\textup{TF}})\cap\mathcal{V}_\ell,
\]
we find straightforwardly that $\Vert\,\phi_{\ell,n}\,\Vert=1$ (since we take $\Vert\,Y_\ell^0\,\Vert=1$) and
\begin{equation}\label{WeylSequence}
\Vert\,\widetilde{H}_{Z_\ell}^{\text{TF}}\phi_{\ell,n}-\mu_\ell\phi_{\ell,n}\,\Vert=\Vert\,H_{Z_\ell,\ell}^{\text{TF}}\psi_{\ell,n}-\mu_\ell\psi_{\ell,n}\,\Vert\longrightarrow0
\end{equation}
as $n\to\infty$. By taking a subsequence (still denoted $\{\phi_{\ell,n}\}_{n=1}^{\infty}$) we can assume that the $\phi_{\ell,n}$'s converge towards some $\phi_\ell$ weakly in $L^2(\mathbb{R}_+)\otimes\operatorname{span}Y_\ell^0$ as $n\to\infty$. Now this $\phi_\ell$ cannot be $0$ as then we would have $\mu_\ell\in\sigma_{\operatorname{ess}}(\widetilde{H}_{Z_\ell}^{\text{TF}})\cap(-\infty,0)=\emptyset$. Moreover, this weak convergence together with (\ref{WeylSequence}) yields
\[
\langle\,\phi_\ell\,,\widetilde{H}_{Z_\ell}^{\text{TF}}\phi\,\,\rangle=\lim_{n\to\infty}\langle\,\phi_{\ell,n}\,,\widetilde{H}_{Z_\ell}^{\text{TF}}\phi\,\,\rangle=\lim_{n\to\infty}\langle\,\widetilde{H}_{Z_\ell}^{\text{TF}}\phi_{\ell,n}\,,\phi\,\,\rangle=\langle\,\mu_\ell\phi_\ell\,,\phi\,\,\rangle
\]
for any $\phi\in D(\widetilde{H}_{Z_\ell}^{\text{TF}})$ proving $\phi_\ell\in D((\widetilde{H}_{Z_\ell}^{\text{TF}})^*)=D(\widetilde{H}_{Z_\ell}^{\text{TF}})$ and $\widetilde{H}_{Z_\ell}^{\text{TF}}\phi_\ell=\mu_\ell\phi_\ell$. This finishes the proof.
\qed
Consider now the sequences $\{Z_\ell\}_{\ell=1}^{\infty}$ and $\{\phi_\ell\}_{\ell=1}^{\infty}$ from Lemmas \ref{NoNormLemma}($c$) and \ref{EigenVectors} respectively. As in the proof of Lemma \ref{EasyImplication} we take a subsequence $\{Z_{\ell_k}\}_{k=1}^{\infty}$ of $\{Z_\ell\}_{\ell=1}^{\infty}$ so that $Z_{\ell_k}\to\infty$ and (\ref{MainResultEquationTF}) as $k\to\infty$ for some $\tau$. We claim that $\{H_{Z_{\ell_k}}^{\text{TF}}\}_{k=1}^{\infty}$ is an example of a sequence of Thomas-Fermi atoms which converges in the strong resolvent sense -- this follows directly from Theorem \ref{MainResultTF} -- but \emph{\textbf{not}} in the norm resolvent sense.

To prove the last part of this statement we choose natural numbers $p_1<p_2<p_3<\cdots$ so that
\begin{equation}\label{NonNegativeComponent}
\frac{\ell_{p_k}(\ell_{p_k}+1)}{x^2}-\Phi_{Z_{\ell_k}}^{\text{TF}}(x)\geq0,\qquad\text{implying}\qquad\widetilde{H}_{Z_{\ell_k}}^{\text{TF}}\big\vert_{D(\widetilde{H}_{Z_{\ell_k}}^{\textup{TF}})\cap\mathcal{V}_{\ell_{p_k}}}\geq0,
\end{equation}
where the $\mathcal{V}_{\ell_{p_k}}$'s are as in Lemma \ref{EigenVectors}. By the construction we have of course $p_k\geq k$. If we let $h$ be a smooth function with compact support in $(-\infty,0)$ we claim that $h(\widetilde{H}_{Z_{\ell_k}}^{\text{TF}})\phi_{\ell_{p_k}}=0$. Indeed, $\mathcal{V}_{\ell_{p_k}}$ is an invariant subspace for $\widetilde{H}_{Z_{\ell_k}}^{0,\text{TF}}$ and hence also for $\widetilde{H}_{Z_{\ell_k}}^{\text{TF}}$, and since $\phi_{\ell_{p_k}}\in\mathcal{V}_{\ell_{p_k}}$ this assertion follows from (\ref{NonNegativeComponent}) and the abstract functional calculus. As additionally $h(\widetilde{H}_{Z_{\ell_{p_k}}}^{\text{TF}})\phi_{\ell_{p_k}}=h(\mu_{\ell_{p_k}})\phi_{\ell_{p_k}}$ we can choose $h$ to be $1$ on a neighbourhood around $-1$ and obtain
\[
\bigl\Vert\,\phi_{\ell_{p_k}}\,\bigr\Vert=\bigl\Vert\,h(\widetilde{H}_{Z_{\ell_{p_k}}}^{\text{TF}})\phi_{\ell_{p_k}}-h(\widetilde{H}_{Z_{\ell_k}}^{\text{TF}})\phi_{\ell_{p_k}}\,\bigr\Vert\leq\bigl\Vert\,\phi_{\ell_{p_k}}\,\bigr\Vert\cdot\bigl\Vert\,h(\widetilde{H}_{Z_{\ell_{p_k}}}^{\text{TF}})-h(\widetilde{H}_{Z_{\ell_k}}^{\text{TF}})\,\bigr\Vert
\]
for sufficiently large $k$ where the last norm is the usual operator norm. This proves that $\{H_{Z_{\ell_k}}^{\text{TF}}\}_{k=1}^{\infty}$ simply cannot be convergent in the norm resolvent sense.

\section*{Acknowledgements}
This work was supported in parts by the VILLUM Foundation grant no. 10059. We thank Jan Derezinski for valuable discussions and input. Part of this work was done while AB was visiting the Mathematics Department at LMU Munich. AB thanks Prof.\ P.T.~Nam for his hospitality while visiting there. 

\printbibliography

\end{document}